\documentclass[%
 reprint,
superscriptaddress,
 amsmath,amssymb,
 aps,
]{revtex4-2}
\usepackage{siunitx}
\usepackage{graphicx}  
\usepackage{subfigure}
\usepackage{multirow}
\usepackage{tabularx}
\usepackage{fancyhdr}
\usepackage{longtable}
\usepackage{parskip}
\usepackage[T1]{fontenc}
\usepackage{dcolumn}   
\usepackage{gensymb}
\usepackage{bm}        
\usepackage{amsfonts}  
\usepackage{amsmath}   
\usepackage{amssymb}   
\usepackage{hhline}
\usepackage{xcolor}
\usepackage{ulem}
\usepackage{gensymb}
\usepackage{soul}
\begin{document}

\title{Magnetic properties of binary alloys Ni$_{1-x}$Mo$_{x}$ and Ni$_{1-y}$Cu$_{y}$ close to critical concentrations}

\author{R.-Z. Lin}
\altaffiliation{These authors contributed equally to this work}
\affiliation {\it Department of Physics, National Cheng Kung University, Tainan 701, Taiwan}
\affiliation {\it Center for Quantum Frontiers of Research \& Technology (QFort), National Cheng Kung University, Tainan 701, Taiwan}
\affiliation{\it Taiwan Consortium of Emergent Crystalline Materials, National Science and Technology Council, Taipei 10622, Taiwan}
\author{J.-X. Hsu}
\altaffiliation{These authors contributed equally to this work}
\affiliation {\it Department of Physics, National Cheng Kung University, Tainan 701, Taiwan}
\affiliation {\it Center for Quantum Frontiers of Research \& Technology (QFort), National Cheng Kung University, Tainan 701, Taiwan}
\affiliation{\it Taiwan Consortium of Emergent Crystalline Materials, National Science and Technology Council, Taipei 10622, Taiwan}
\author{E.-P. Liu}
\affiliation {\it Center for Condensed Matter Sciences, National Taiwan University, Taipei 10617, Taiwan}
\affiliation {\it Department of Physics, National Taiwan University, Taipei 10617, Taiwan}
\author{W.-T. Chen}
\affiliation {\it Center for Condensed Matter Sciences, National Taiwan University, Taipei 10617, Taiwan}
\affiliation {\it Center of Atomic Initiative for New Materials, National Taiwan University, Taipei 10617, Taiwan}
\affiliation{\it Taiwan Consortium of Emergent Crystalline Materials, National Science and Technology Council, Taipei 10622, Taiwan}
\author{C.-L. Huang}
\email{clh@phys.ncku.edu.tw}
\affiliation {\it Department of Physics, National Cheng Kung University, Tainan 701, Taiwan}
\affiliation {\it Center for Quantum Frontiers of Research \& Technology (QFort), National Cheng Kung University, Tainan 701, Taiwan}
\affiliation{\it Taiwan Consortium of Emergent Crystalline Materials, National Science and Technology Council, Taipei 10622, Taiwan}

\date{\today}

\begin{abstract}  
The search for the ferromagnetic quantum critical point (FM QCP) has always been a captivating research topic in the scientific community. In pursuit of this goal, we introduced nonmagnetic transition metals to alloy with elemental nickel, and studied the magnetic properties of nickel binary alloys Ni$_{1-x}$Mo$_{x}$ and Ni$_{1-y}$Cu$_{y}$ as a function of $x$ and $y$ up to the critical concentrations $x_{cr}$ and $y_{cr}$ at which the FM transition $T_{\rm C}$ disappears. $T_{\rm C}-x(y)$ phase diagrams were constructed via the Arrott-Noakes scaling of magnetization data. An enhanced Sommerfeld coefficient (the value of $C/T$ as $T \rightarrow 0$) is observed near $y_{cr}$, manifesting the effect of quantum fluctuations near the quantum phase transition. It is evident that $C/T$ diverges with $-$log$T$ down to 0.1~K in the vicinity of $y_{cr}$, suggests the plausible FM QCP in Ni$_{1-y}$Cu$_{y}$. However, in the case of Ni$_{1-x}$Mo$_{x}$, although the enhancement of the Sommerfeld coefficient is also observed near $x_{cr}$, the spin glass behavior is identified through the ac magnetic susceptibility measurement. This observation rules out the possibility of the existence of the FM QCP in Ni$_{1-x}$Mo$_{x}$.

\end{abstract}

\maketitle
\section{INTRODUCTION}
A ferromagnetic quantum critical point (FM QCP) is one of the most exotic quantum states in condensed matter physics. Such a state has a FM long-range ordered ground state before tuning. When some temperature-irrelevant tuning parameter, such as, physical pressure or chemical substitution, is applied, the transition temperature $T_{\rm C}$ can be continuously suppressed to absolute zero at which the FM order is destroyed by quantum fluctuations of the order parameter \cite{Brando2016}. The increasing attention drawn to search for the FM QCP is not only due to its rarity compared the antiferromagnetic counterpart but also because the quantum fluctuations near the QCP often lead to emergent physics phenomena, e.g., unconventional superconductivity and non-Fermi-liquid (NFL) behavior \cite{Stewart2001,Loehneysen2007,Brando2016}. 

Tuning a ferromagnet to the QCP is not trivial \cite{Brando2016}. The only controlled method that has been experimentally \cite{Huang2016,Goko2017,Sales2017,Lai2018,Huang2020} and theoretically \cite{Belitz1999,Sang2014,Kirkpatrick2014,Kirkpatrick2015} verified involves introducing an appropriate amount of chemical disorder. To meet this condition, a binary alloy of a nonmagnetic transition metal $A$ and an FM element, such as Fe, Co, or Ni, is arguably the simplest one. It is natural to expect that the $T_{\rm C}$ of the binary FM alloy can be suppressed as the concentration of $A$ increases. This is indeed observed in several Ni$_{1-x}A_{x}$ alloys \cite{Gupta1964,Brinkman1968,Boelling1968,Gregory1975,Nicklas1999,Ubaid-Kassis2010}. However, for all the Ni$_{1-x}A_{x}$ alloys, only one FM QCP exists: Ni$_{1-x}$Rh$_{x}$ with $x = 0.375$ where NFL behavior is reported \cite{Huang2020,Lin2022}. Therefore, it is worth to further investigate the magnetic properties of other Ni$_{1-x}A_{x}$ alloys to explore other possible FM QCPs. In this work, we report a detailed study on Ni$_{1-x}$Mo$_{x}$ and Ni$_{1-y}$Cu$_{y}$ binary alloys. Structure analysis confirmed the face-centered-cubic structure of nickel in all samples. The degree of chemical disorder can be identified through electrical resistivity measurements. The $T_{\rm C}-x(y)$ phase diagrams were constructed with the aid of magnetic susceptibility measurements and scaling analysis on isothermal magnetization curves. Enhanced low-temperature specific heat $C$ was observed near the critical concentrations ($x_{cr}$ and $y_{cr}$) of both alloys at which the $T_{\rm C}$ vanishes, indicating the effect of quantum fluctuations. Such the effect extends down to $T = 0.1$~K in Ni$_{1-y}$Cu$_{y}$, where $C/T$ varies with $-$log$T$ in the vicinity of $y_{cr}$, suggesting the existence of a plausible FM QCP. Finally, the ac magnetic susceptibility measurement on Ni$_{1-x}$Mo$_{x}$ at $x_{cr}$ revealed spin glass behavior, most probably due to the random alignment of magnetic moments in the system. This result rules out the existence of the FM QCP in the Ni$_{1-x}$Mo$_{x}$ alloy.

\section{RESEARCH METHODS}
Polycrystalline Ni$_{1-x}$Mo$_{x}$ and Ni$_{1-y}$Cu$_{y}$ samples were prepared by arcmelting high-purity Ni, Mo, and Cu elements. Samples were sealed under vacuum within a quartz tube. Subsequently, Ni$_{1-x}$Mo$_{x}$ underwent annealing at 1,000$^\circ$~C for a duration of five days, and was cooled in air. Ni$_{1-y}$Cu$_{y}$ was subjected to annealing at 700$^\circ$~C for a week, and was quenched in water to avoid the phase separation below 400$^\circ$~C \cite{Turchanin2007}. X-ray measurements were carried out using a Bruker D2 Phaser diffractometer. The chemical composition was determined by electron probe microanalysis (EPMA) using a JEOL JXA 8530F Hyperprobe. Resistivity and specific heat were measured using a Quantum Design (QD) Dynacool physical property measurement system (PPMS). For specific heat, the contribution of addenda (grease and the platform) was measured before the measurements of each sample. The specific heat down to 0.06~K was measured using the QD Dynacool PPMS equipped with a $^{3}$He/$^{4}$He dilution refrigerator. The electrical resistivity was measured with the standard four-probe method. Magnetization and magnetic susceptibility measurements were carried out using a QD magnetic property measurement system.

\begin{figure}
\includegraphics[width=\columnwidth]{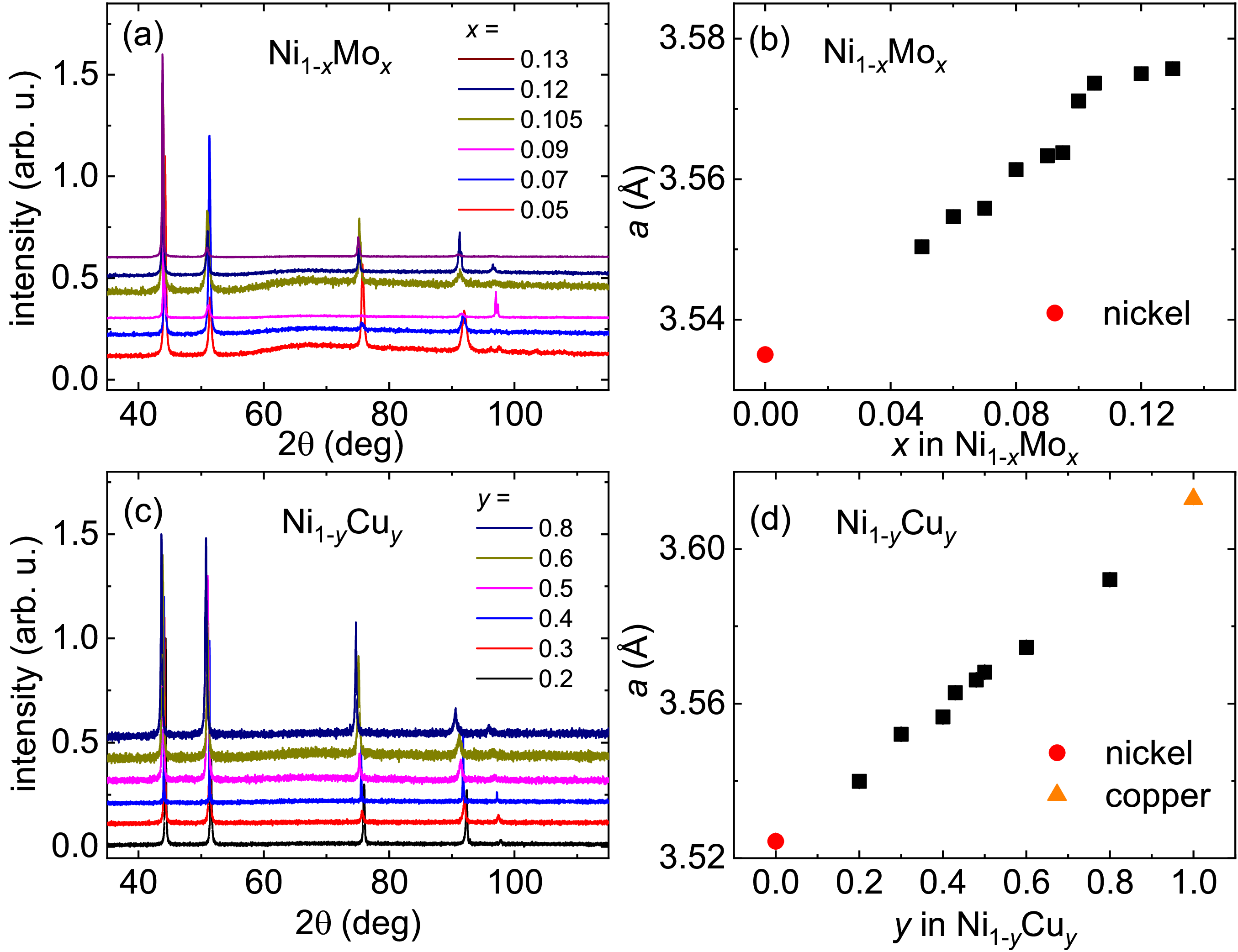}
\caption{(a)(c) Representative x-ray diffraction patterns of Ni$_{1-x}$Mo$_{x}$ and Ni$_{1-y}$Cu$_{y}$ at room temperature. (b)(d) Refined lattice constants as a function of Mo and Cu concentrations.}  
\label{xray} 
\end{figure}

The phase purity and the face-centered-cubic structure of samples was confirmed by room temperature powder x-ray diffraction measurements, as shown in Fig.~\ref{xray}(a) and (c). For the parent metals, both Ni and Cu crystallized in FCC structure with $Fm\bar{3}m$ space group, while Mo metal adapts BCC structure with $Im\bar{3}m$ space group. Due to th nature of the parent metals, the resulting doping limit of Ni$_{1-x}$Mo$_{x}$ samples is rather low, and full doping was achieved in Ni$_{1-y}$Cu$_{y}$ samples. Because of the extreme hardness of the arcmelted samples, x-ray diffraction data were collected on cut and polished specimens. The x-ray patterns were refined using GSAS II. The resulting lattice parameters vary linearly with nominal Mo and Cu concentrations, as shown in Fig.~\ref{xray}(b) and (d), consistent with the Vegard's law.  The EPMA determined compositions also vary linearly with the nominal composition values, as shown in Fig.~\ref{EPMA}. Therefore, nominal composition is adopted throughout the main text. 

\section{RESULTS and DISCUSSIONS}
\subsection{Electrical resistivity}
\begin{figure}
\includegraphics[width=\columnwidth]{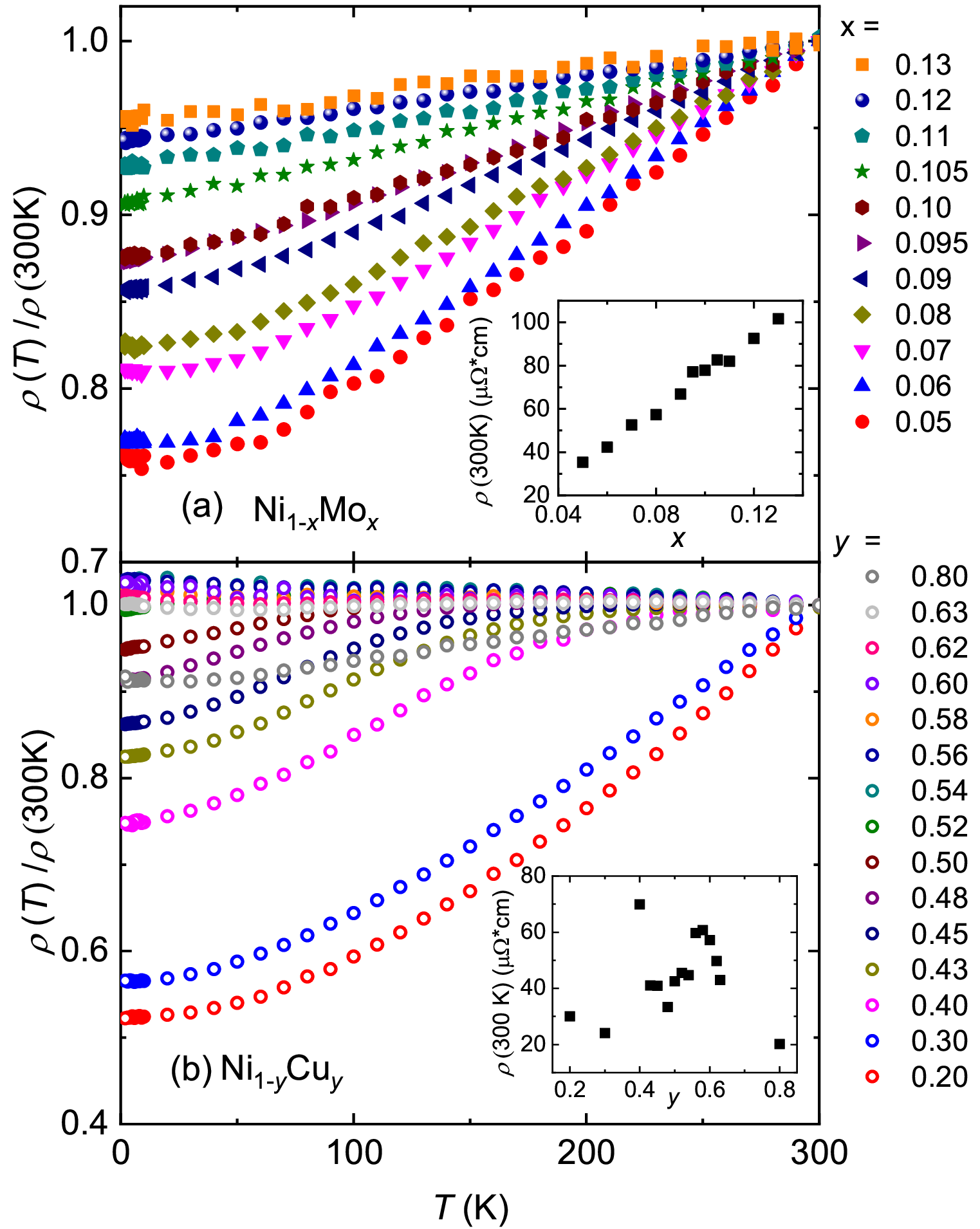}
\caption{Normalized resistivity $\rho/\rho_{\rm{300 K}}$ for (a) Ni$_{1-x}$Mo$_{x}$ and (b) Ni$_{1-y}$Cu$_{y}$. The insets show the absolute value of $\rho$ at 300~K for each concentration.}  
\label{rho} 
\end{figure}

 Figure~\ref{rho} shows the temperature dependence of resistivity normalized to its value at 300~K, $\rho(T)/\rho_{\rm{300 K}}$. For Ni$_{1-x}$Mo$_{x}$, $\rho(T)/\rho_{\rm{300 K}}$ decreases as the temperature decreases, as expected for a metallic sample. The $\rho_{\rm{300 K}}$ value shows a monotonic increase with $x$. The fact that $\rho(T)/\rho_{\rm{300 K}}$ at 2~K and $\rho_{\rm{300 K}}$ are both the smallest for $x = 0.05$ (Fig.~\ref{rho}(a) and the inset), the largest for $x = 0.13$, and increase with $x$, indicates that the disorder effect increases with the amount of Mo in Ni$_{1-x}$Mo$_{x}$. 

\begin{figure}
\includegraphics[width=\columnwidth]{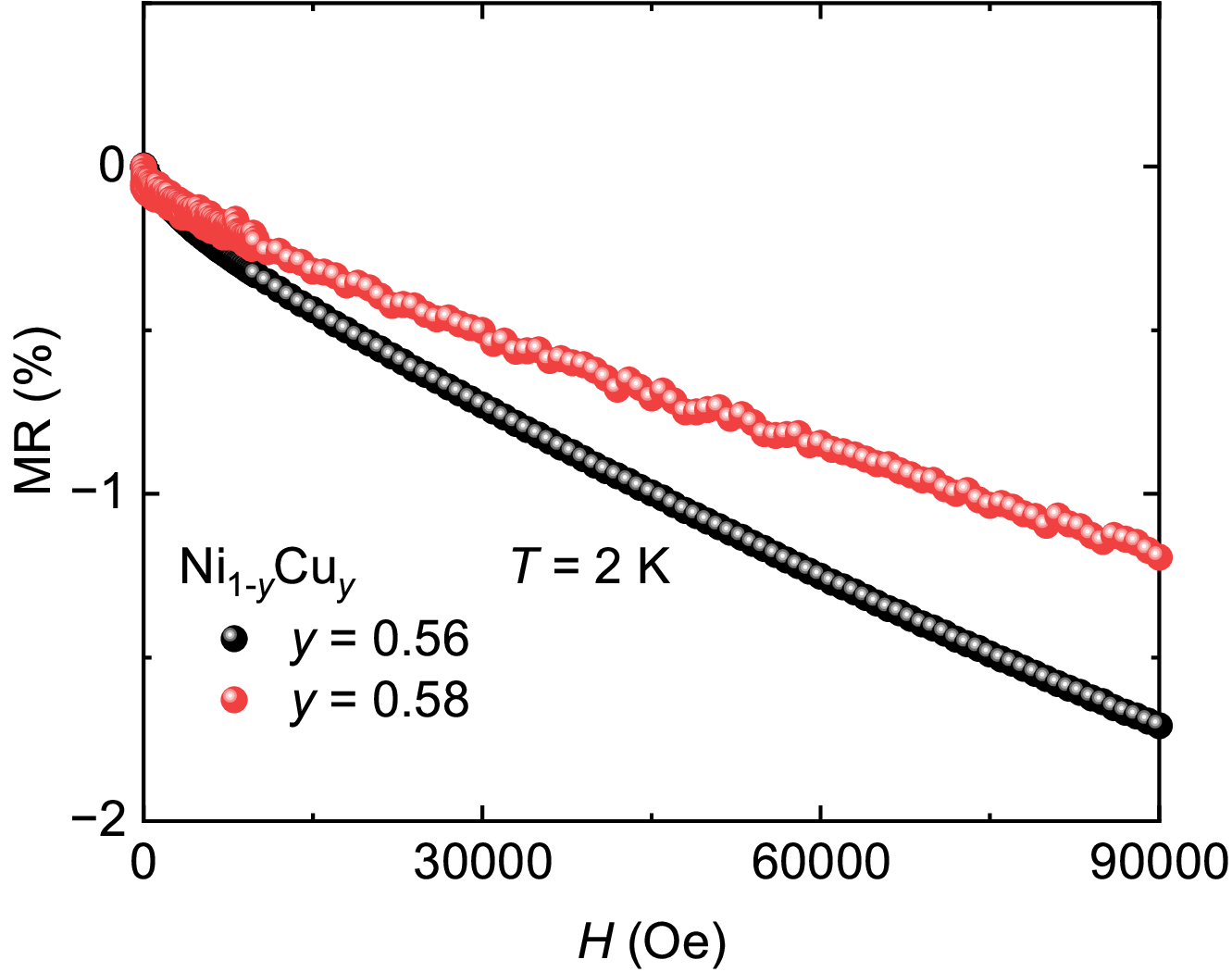}
\caption{Magnetoresistance MR(\%) = ($\rho(H)-\rho(0))/\rho(0)\times 100$\% for Ni$_{1-y}$Cu$_{y}$ measured at 2~K.}  
\label{MR} 
\end{figure}

In contrast with the low doping in Ni$_{1-x}$Mo$_{x}$, much higher disorder effect is registered for Ni$_{1-y}$Cu$_{y}$ with high doping. This is also reflected in a larger error bar of EPMA results in Ni$_{1-y}$Cu$_{y}$ as compared to Ni$_{1-x}$Mo$_{x}$ (see Fig.~\ref{EPMA}). The metallic behavior, i.e., $\rho(T)/\rho_{\rm{300 K}}$ decreases upon cooling, is only seen in $y$ = 0.20-0.50, as shown in Fig.~\ref{rho}(b). For $0.80 > y \geq 0.52$, $\rho(T)/\rho_{\rm{300 K}}$ slightly increases as the temperature decreases from 300 to 2~K. This might be due to structural disorder as a result of Cu substitution. A small negative transverse magnetoresistance in $y = 0.56$ and 0.58 samples are considered evidence of charge transport involving weak Anderson localization, as shown in Fig.~\ref{MR} \cite{Diaz2021}. For $y = 0.80$, $\rho(T)/\rho_{\rm{300 K}}$ again shows metallic behavior with a much smaller rate of $d\rho/dT$ as compared to $y$ = 0.20-0.50 samples, which can be viewed as an effect of site disorder.

\subsection{Magnetic susceptibility}
\begin{figure}
\includegraphics[width=\columnwidth]{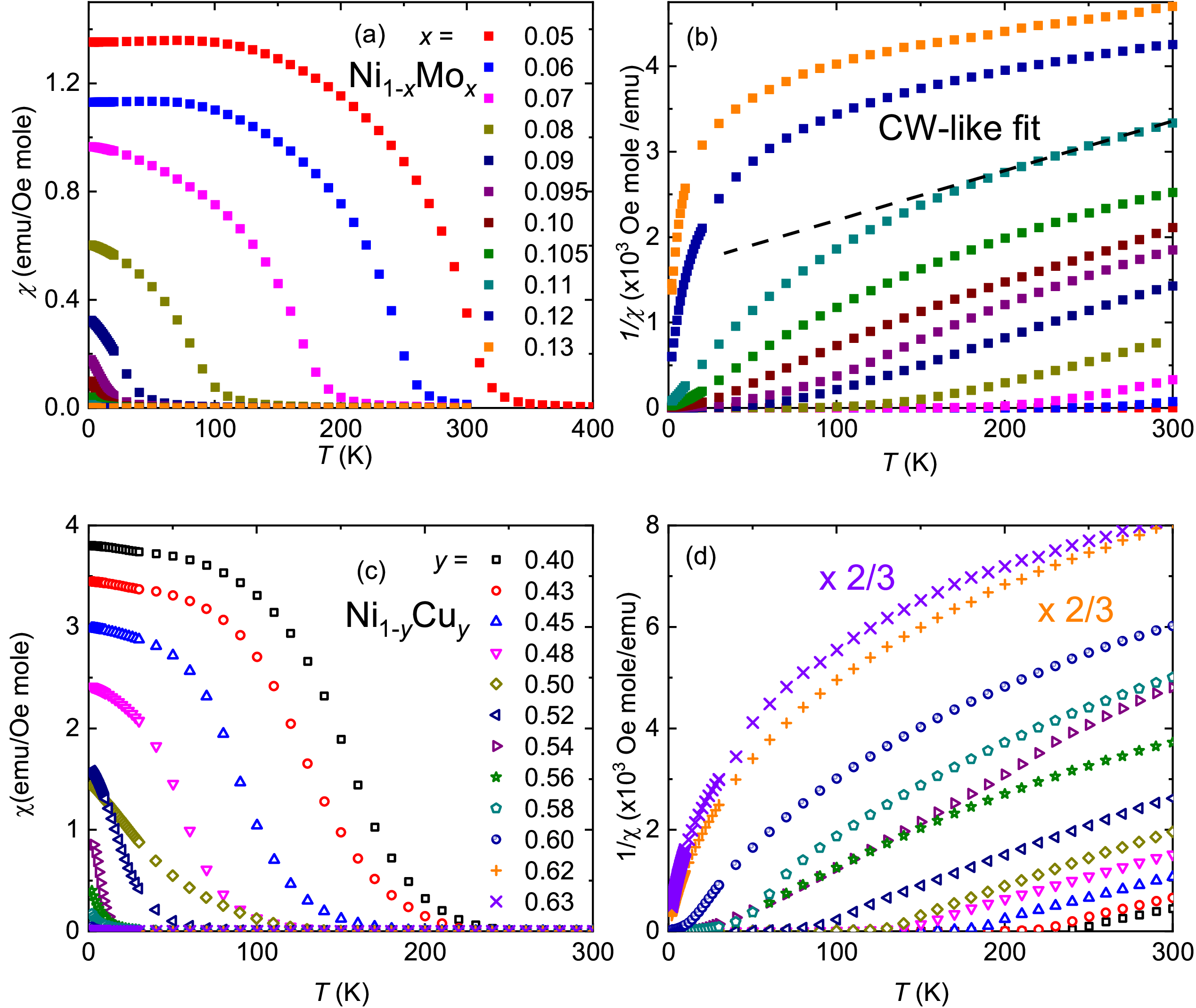}
\caption{(a)(c) Magnetic susceptibility $\chi$ of Ni$_{1-x}$Mo$_{x}$ and Ni$_{1-y}$Cu$_{y}$. (b)(d) Inverse $\chi$ vs. $T$. (a) and (b); (c) and (d) use the same legends.}  
\label{chiT} 
\end{figure}

The field-cooled magnetic susceptibility $\chi = M/H$ under the magnetic field $H = 1000$ Oe of Ni$_{1-x}$Mo$_{x}$ and under the magnetic field $H = 100$ Oe of Ni$_{1-y}$Cu$_{y}$ is shown in Fig.~\ref{chiT}(a) and (c), respectively. When $T$ is lowered through a paramagnetic-to-FM transition, $\chi$ sharply increases. The $T_{\rm C}$ decreases with increasing $x$ and $y$, as expected in diluted magnetic alloys, and is consistent with the previous report \cite{Amamou1975}. 

\begin{figure}
\includegraphics[width=\columnwidth]{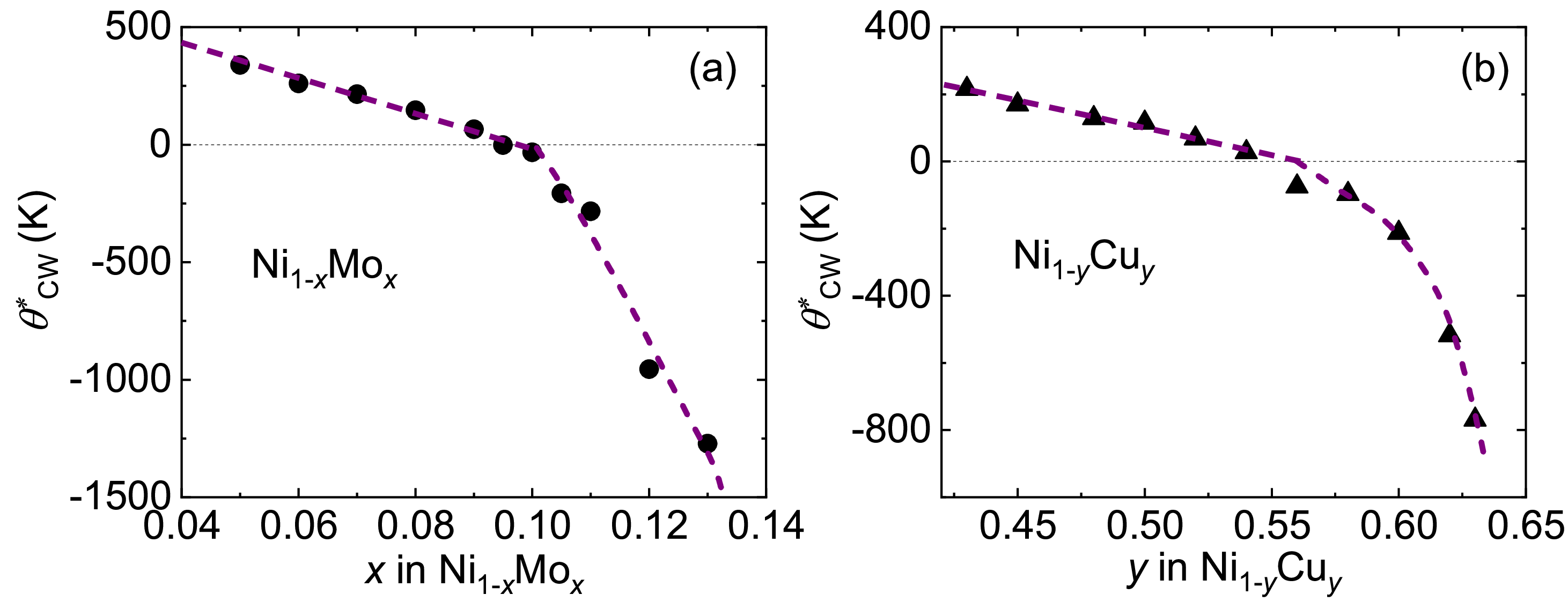}
\caption{The Curie-Weiss-like temperature $\theta^*_{\rm CW}$ as a function of (a) $x$ in Ni$_{1-x}$Mo$_{x}$ and (b) $y$ in Ni$_{1-y}$Cu$_{y}$. Dashed lines are guides to the eye.}  
\label{CW} 
\end{figure}

For itinerant magnetic systems, the temperature dependence of the inverse magnetic susceptibility is analogous to the Curie-Weiss law developed for local moment systems, i.e.,
\begin{equation}
   \chi (T)= \frac{D}{T-\theta^*_{\rm CW}},
   \label{CWlike}
\end{equation}
where $D$ is a constant and $\theta^*_{\rm CW}$ is Curie-Weiss-like (CW-like) temperature \cite{Santiago2017,Mugiraneza2022}. In other itinerant ferromagnets as the magnetic order is tuned by chemical substitution, the positive $\theta^*_{\rm CW}$ has been found to trace $T_{\rm C}$ well, decrease with increasing substitutions, and change sign from positive to negative around the critical concentration \cite{Fuchs2014}. We fit the high temperature $\chi^{-1}(T)$ using Eq.~(\ref{CWlike}), as shown by a dashed line in Fig.~\ref{chiT}(b), and obtained $\theta^*_{\rm CW}$ vs. $x$ and $y$ in Fig.~\ref{CW}(a) and (b), respectively. The data shows that $\theta^*_{\rm CW}$ changes sign at $x = 0.10$ and $y = 0.56$. Both values are very close to the critical concentrations $x_{cr}$ and $y_{cr}$ determined from the $T_{\rm C}$-$x(y)$ phase diagrams, which will be discussed later. The coincidence of the concentration where the $\theta^*_{\rm CW}$ changes sign and the critical concentration has been observed in other ferromagnets where QCPs exist at the same tuning parameter \cite{Fuchs2014,Huang2015,Sur2023}. 

\subsection{Arrott-Noakes scaling of magnetization}
\begin{figure}
\includegraphics[width=\columnwidth]{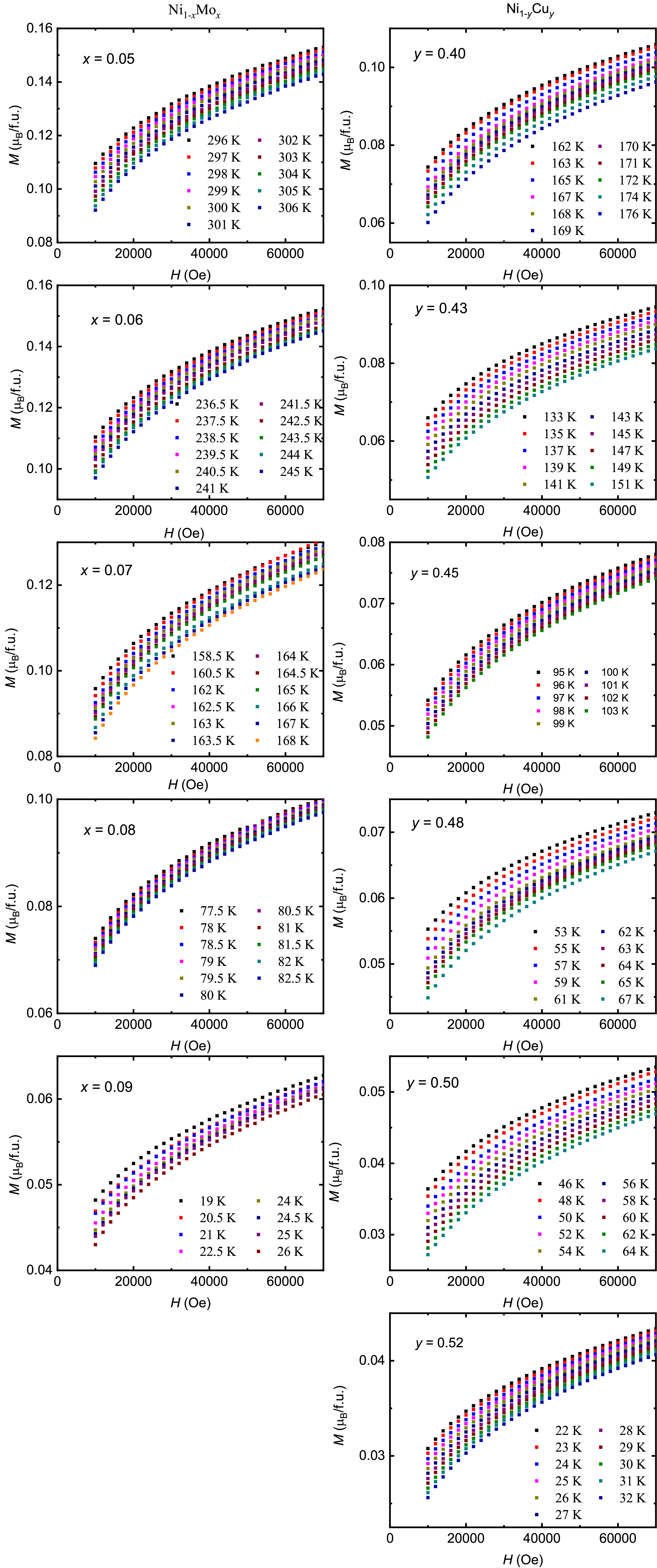}
\caption{Isothermal magnetization curves as a function of the magnetic field for Ni$_{1-x}$Mo$_{x}$ with $x = 0.05-0.09$ (left) and Ni$_{1-y}$Cu$_{y}$ with $y = 0.40-0.52$ (right).}  
\label{MH} 
\end{figure}

\begin{figure}
\includegraphics[width=\columnwidth]{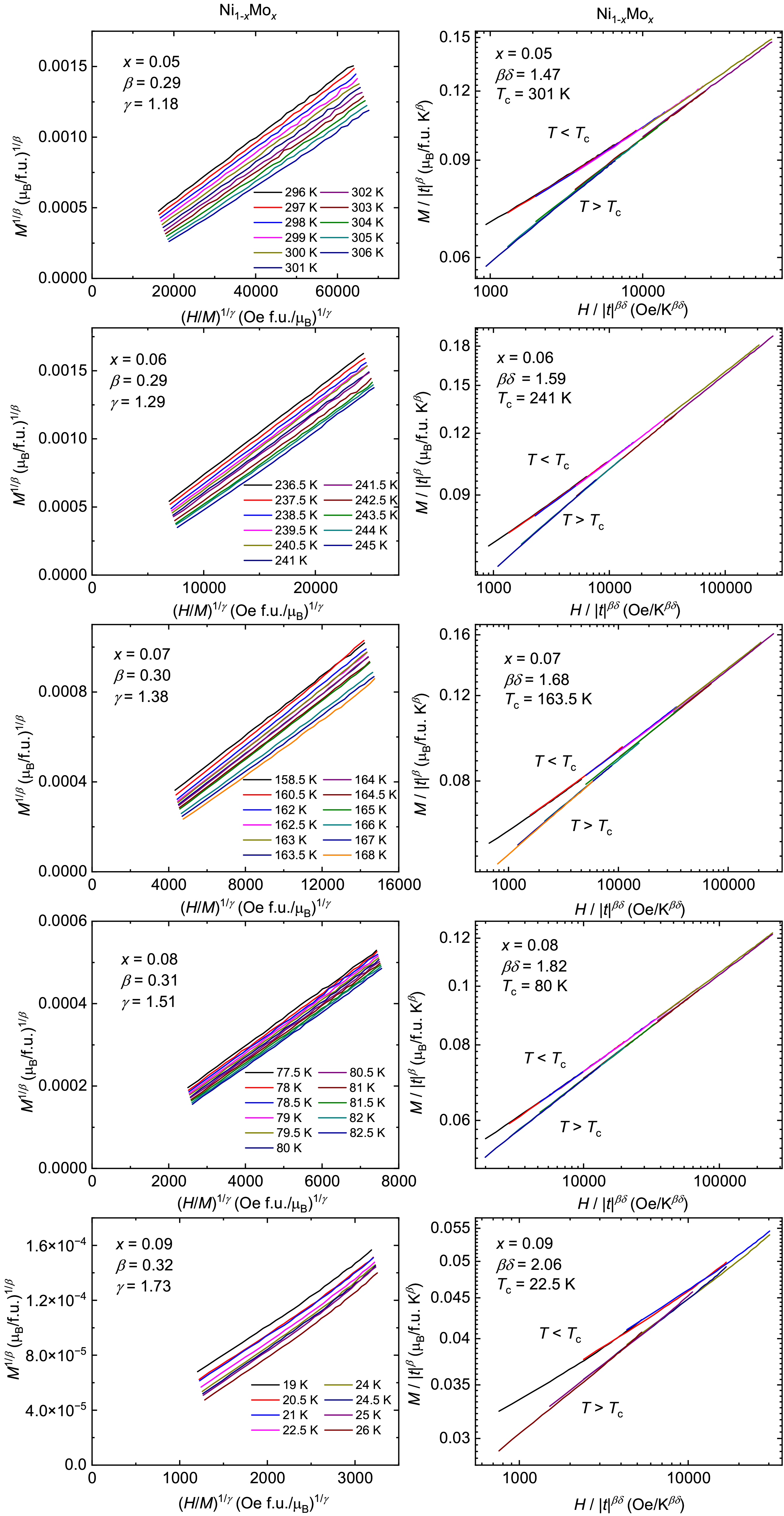}
\caption{The left column shows modified Arrott plots of Ni$_{1-x}$Mo$_{x}$ with $x = 0.05-0.09$. The right column shows Arrott-Noakes scaling plots of the same data displayed in the left column.}  
\label{ArrottNM} 
\end{figure}

\begin{figure}
\includegraphics[width=\columnwidth]{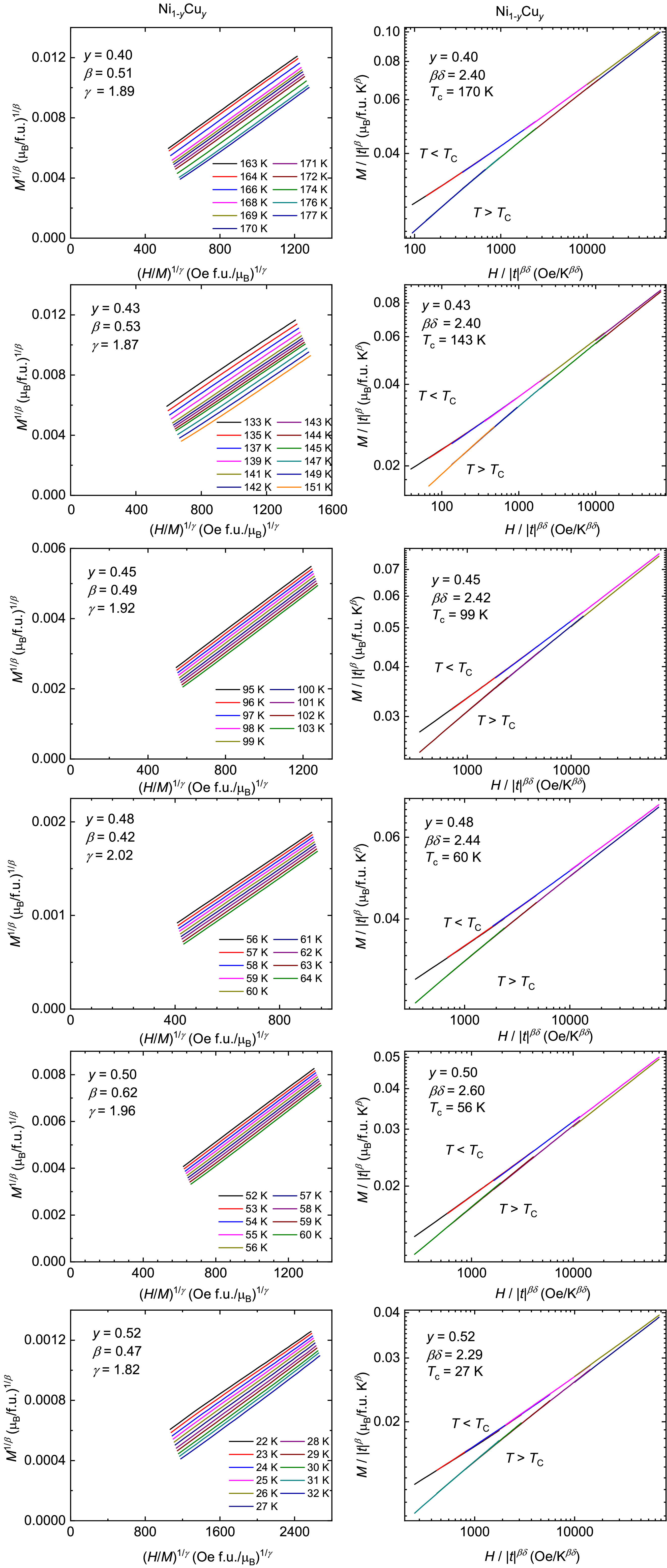}
\caption{The left column shows modified Arrott plots of Ni$_{1-y}$Cu$_{y}$ with $y = 0.40-0.52$. The right column shows Arrott-Noakes scaling plots of the same data displayed in the left column.}  
\label{ArrottNC} 
\end{figure}

In the vicinity of $T_{\rm C}$, the isothermal magnetization curves can be scaled based on an Arrott-Noakes equation of state, from which critical scaling exponents $\beta, \gamma,$ and $\delta$ are extracted \cite{Arrott1967}. The normal isothermal magnetization curves as a function of the magnetic field is shown in Fig.~\ref{MH}. The Arrott-Noakes equation of state, i.e.,
\begin{equation}
    M^{1/\beta}=a^{*}+b^{*}(\frac{H}{M})^{1/\gamma}
    \label{AN}
\end{equation}
leads to the scaling of $M$ since the critical fluctuations are the dominant energy scale. $a^{*}$ and $b^{*}$ are scaling constants. Equation~(\ref{AN}) yields parallel lines, forming the so-called modified Arrott plot, with $M \sim H^{1/\delta}$ at $T_{\rm C}$, and giving rise to the critical exponents $\beta$, $\gamma$, and $\delta$. To ensure the analysis is within the regime where critical fluctuations dominate, magnetic isotherms were measured in evenly spaced temperature steps in the range $(1-3\%)T_{\rm C}\leq T\leq (1+3\%)T_{\rm C}$, where $T_{\rm C}$ corresponds to the temperature at which the modified Arrott plot intersects at the origin. The left columns of Figs.~\ref{ArrottNM} and~\ref{ArrottNC} show the scaling results of Ni$_{1-x}$Mo$_{x}$ and Ni$_{1-y}$Cu$_{y}$, respectively. Critical exponents $\beta$ and $\gamma$ can likewise be determined from the scaling theory which describes the reduced $M$ versus the reduced $\mu_{0}H$ in the form \cite{Kaul1985},
\begin{equation}
    M/|T-T_{\rm {C}}|^{\beta}=f_{\pm}(\mu_{0}H/|T-T_{\rm {C}}|^{\beta+\gamma}),
    \label{Scaling}
\end{equation}
which makes $M$ to fall on two universal curves: one for the FM state, $T < T_{\rm C}$, and the other for the paramagnetic state, $T > T_{\rm C}$. The right columns of Figs.~\ref{ArrottNM} and~\ref{ArrottNC} display the scaling plots for the $M$ shown in the left columns. It is evident that the branches for $T < T_{\rm C}$ and $T > T_{\rm C}$ each collapse onto one curve. The scaling tendencies are apparent, showcasing the congruence between the modified Arrott plot and the Arrott-Noakes scaling analysis.

\begin{figure}
\includegraphics[width=\columnwidth]{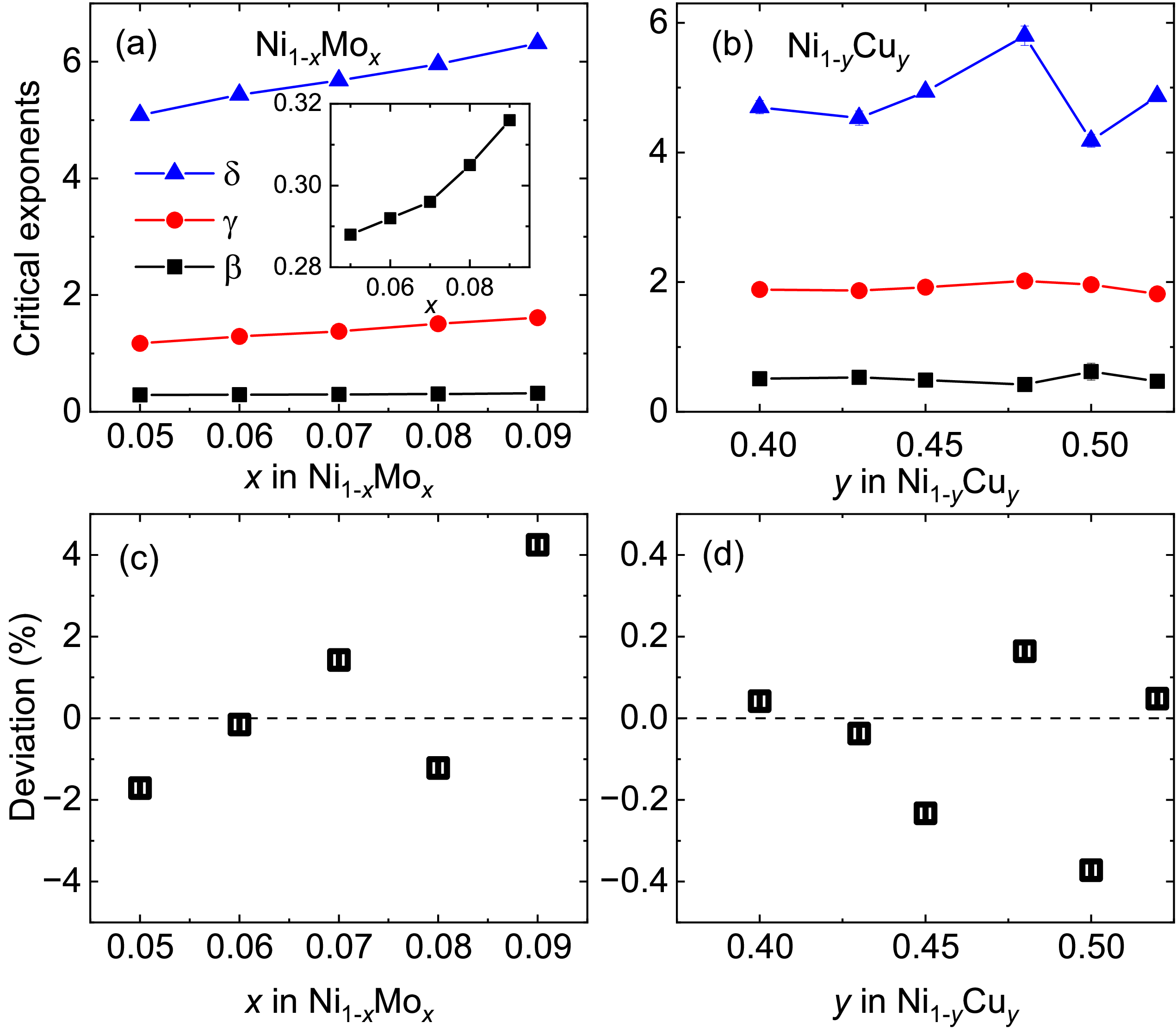}
\caption{Critical exponents as a function of (a) $x$ in Ni$_{1-x}$Mo$_{x}$ and (b) $y$ in Ni$_{1-y}$Cu$_{y}$. Enlarged scale for the exponent $\beta$ is shown in the inset of (a). (c,d) Deviation from the Widom relation. See text for detail.}  
\label{exponent} 
\end{figure}

The results of scaling analysis are summarized in Fig.~\ref{exponent}(a) and (b) where we show the evolution of the critical exponents $\beta$, $\gamma$, and $\delta$ as a function of $x$ and $y$. These exponents are not independent from each other and should follow the Widom relation $\delta = \gamma/\beta + 1$ \cite{Stanley1971}. To test if our scaling results are consistent with the relation, we compare $\delta_{scaling}$ values derived from the scaling analysis with $\delta_{\rm {Widom}}$ values deduced from the Widom relation after introducing $\gamma_{scaling}$ and $\beta_{scaling}$. The deviation = $(\delta_{scaling}-\delta_{\rm {Widom}})/\delta_{\rm {Widom}}\times$ 100\% is shown in Fig.~\ref{exponent}(c) and (d). The deviation is less than 4\% in Ni$_{1-x}$Mo$_{x}$ and 0.4\% in Ni$_{1-y}$Cu$_{y}$. For Ni$_{1-x}$Mo$_{x}$, the critical exponents deviate from the values of pure nickel ($\beta = 0.39$, $\gamma = 1.315$, and $\delta = 4.37$ \cite{Stuesser1986}), and slightly increase as $x$ increases. For Ni$_{1-y}$Cu$_{y}$, the critical exponents do not show a continuous trend as a function of $y$, but fluctuate around the values $\beta \approx 0.51$, $\gamma \approx 1.91$, and $\delta \approx 4.84$. The evolution of $\beta$, $\gamma$, and $\delta$ cannot be explained by currently theoretical models. As the concentration of the nonmagnetic substitution increases, one would expect the magnetic interactions to become more anisotropic compared to pure nickel due to disorder. Subsequently, the critical dimension is reduced so that critical fluctuations become
important. One would expect $\delta$ and $\gamma$ increase, and $\beta$ decreases \cite{Collins1989}. One potential explanation is that quantum fluctuations cause substantial renormalizations of the low-energy theory underlying the classical critical point, and thus drive the system into a strong-coupling regime that is beyond the established theoretical models \cite{Fuchs2014}. 

\begin{figure}
\includegraphics[width=\columnwidth]{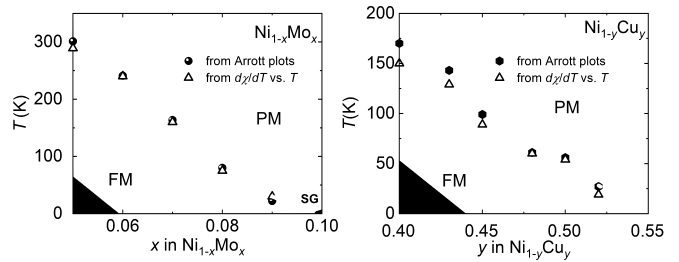}
\caption{Left: $T_{\rm C}-x$ phase diagram of Ni$_{1-x}$Mo$_{x}$. Right: $T_{\rm C}-y$ phase diagram of Ni$_{1-y}$Cu$_{y}$. FM = ferromagnetic state, PM = paramagnetic state, and SG = spin glass.}  
\label{phase} 
\end{figure}

\begin{figure}
\includegraphics[width=\columnwidth]{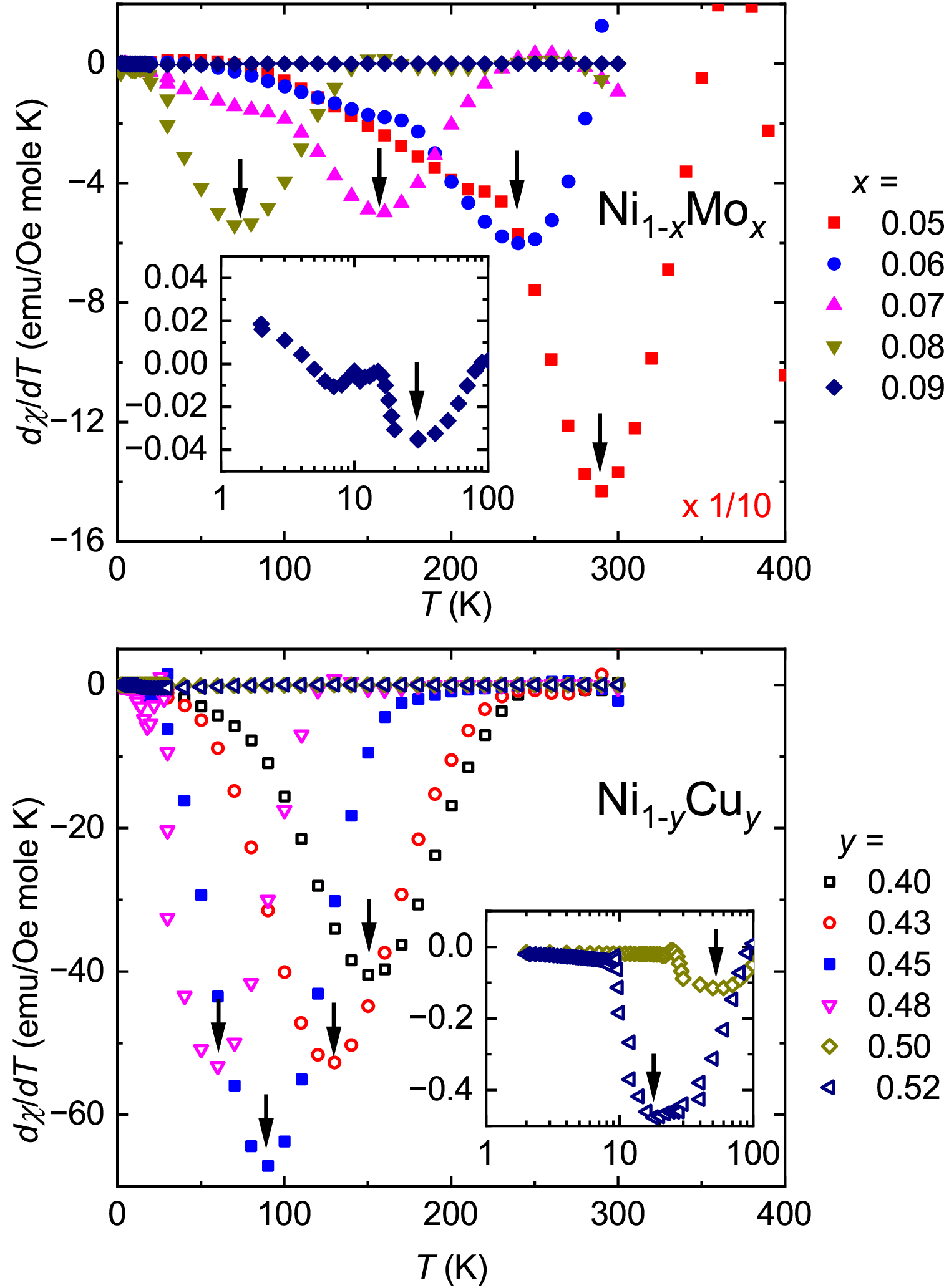}
\caption{$d\chi/dT$ vs. $T$ plots. The arrows mark the $T_{\rm C}$ values. Insets are data with an enlarged scale.}  
\label{dchidT} 
\end{figure}

From modified Arrott plots, we determined $T_{\rm C}$ values for each sample and constructed $T_{\rm C}$ vs. $x$($y$) phase diagrams, as shown in Fig.~\ref{phase}. We also determined $T_{\rm C}$ values from the the maximum negative slope in $d\chi/dT$ vs. $T$ plots (shown in Fig.~\ref{dchidT}), and the results are comparable with the ones from modified Arrott plots. 

\subsection{Low temperature specific heat}

\begin{figure}
\includegraphics[width=\columnwidth]{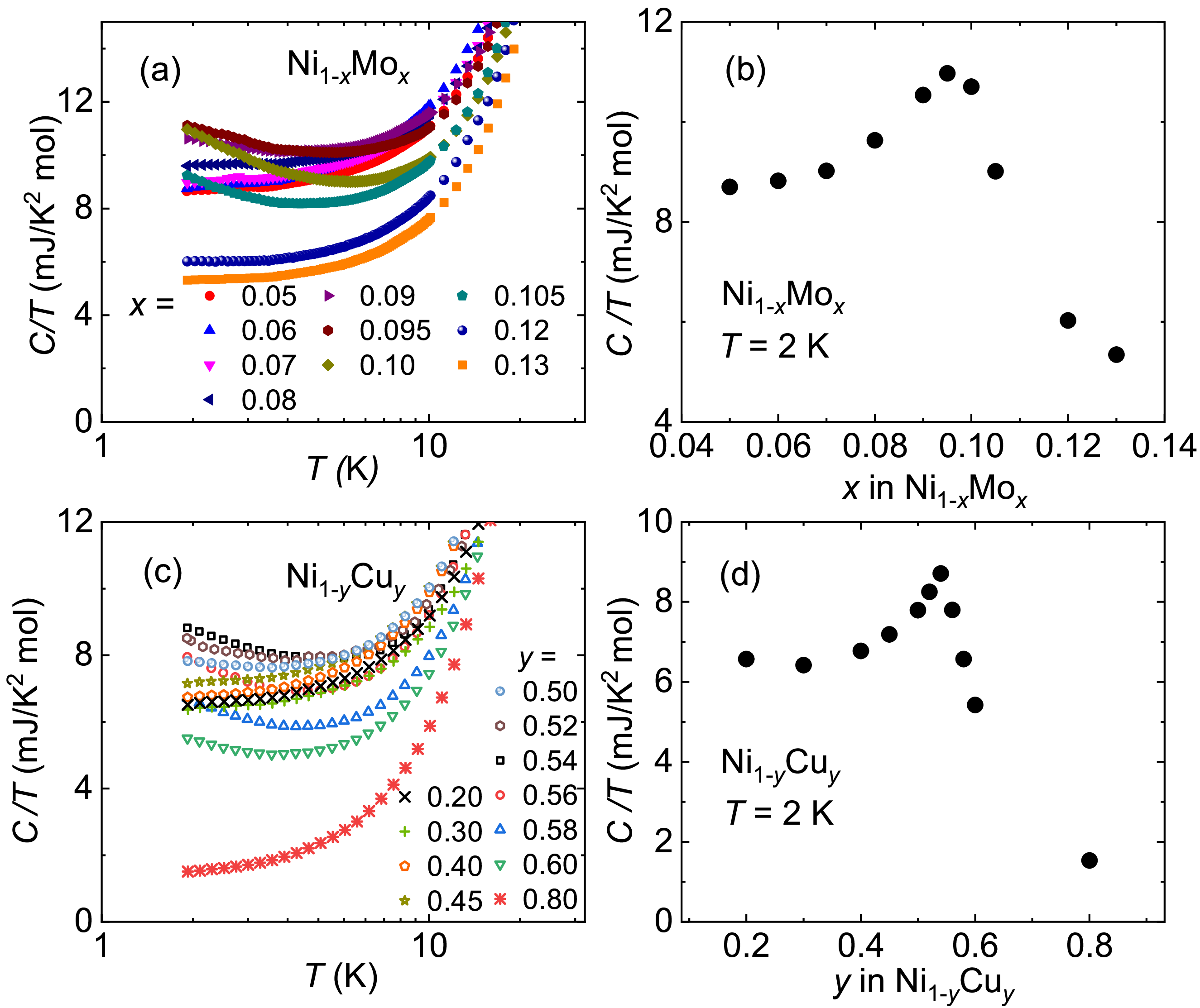}
\caption{Specific heat of (a) Ni$_{1-x}$Mo$_{x}$ and (c) Ni$_{1-y}$Cu$_{y}$ plotted as $C/T$ vs. log$T$. $C/T$ at 2~K as a function of (b) $x$ and (d) $y$.}  
\label{CT} 
\end{figure}

As the $T_{\rm C}$ is suppressed and approaches zero, the influence of thermal fluctuations on the order parameter fluctuations diminishes, while the influence of quantum fluctuations becomes dominant. This is best captured by low-temperature specific heat measurement as the specific heat can directly probe the low-energy magnetic excitations. Figures~\ref{CT}(a) and (c) show the temperature dependence of zero-field specific heat $C/T$ for Ni$_{1-x}$Mo$_{x}$ and Ni$_{1-y}$Cu$_{y}$, respectively. We note that usually no anomaly can be seen at $T_{\rm C}$ in specific heat (and resistivity) measurements of itinerant magnets. This is due to the fact the entropy change associated with the magnetic transition in the itinerant magnets is much smaller than in the local-moment counterparts. For the concentrations significantly below $x_{cr}$ and $y_{cr}$, $C/T$ saturates at constant values as the temperature is decreased to 1.8~K, indicating Fermi-liquid behavior for metals. As $x$ and $y$ approach $x_{cr}$ and $y_{cr}$, $C/T$ elevates and increases as the temperature decreases. We plot the $C/T$ value at 2~K as a function of $x$ and $y$ in Fig.~\ref{CT}(b) and (d), respectively. It is evident that $C/T$ peaks at $x_{cr} \sim 0.095$ and $y_{cr} \sim 0.54$. In several magnetic systems tuned to the QCP, the similar phenomenon has been observed where the maximal $C/T$ centers around the critical concentration, reflecting the effect of enhanced quantum fluctuations \cite{Huy2007,Lai2018}.

\begin{figure}
\includegraphics[width=\columnwidth]{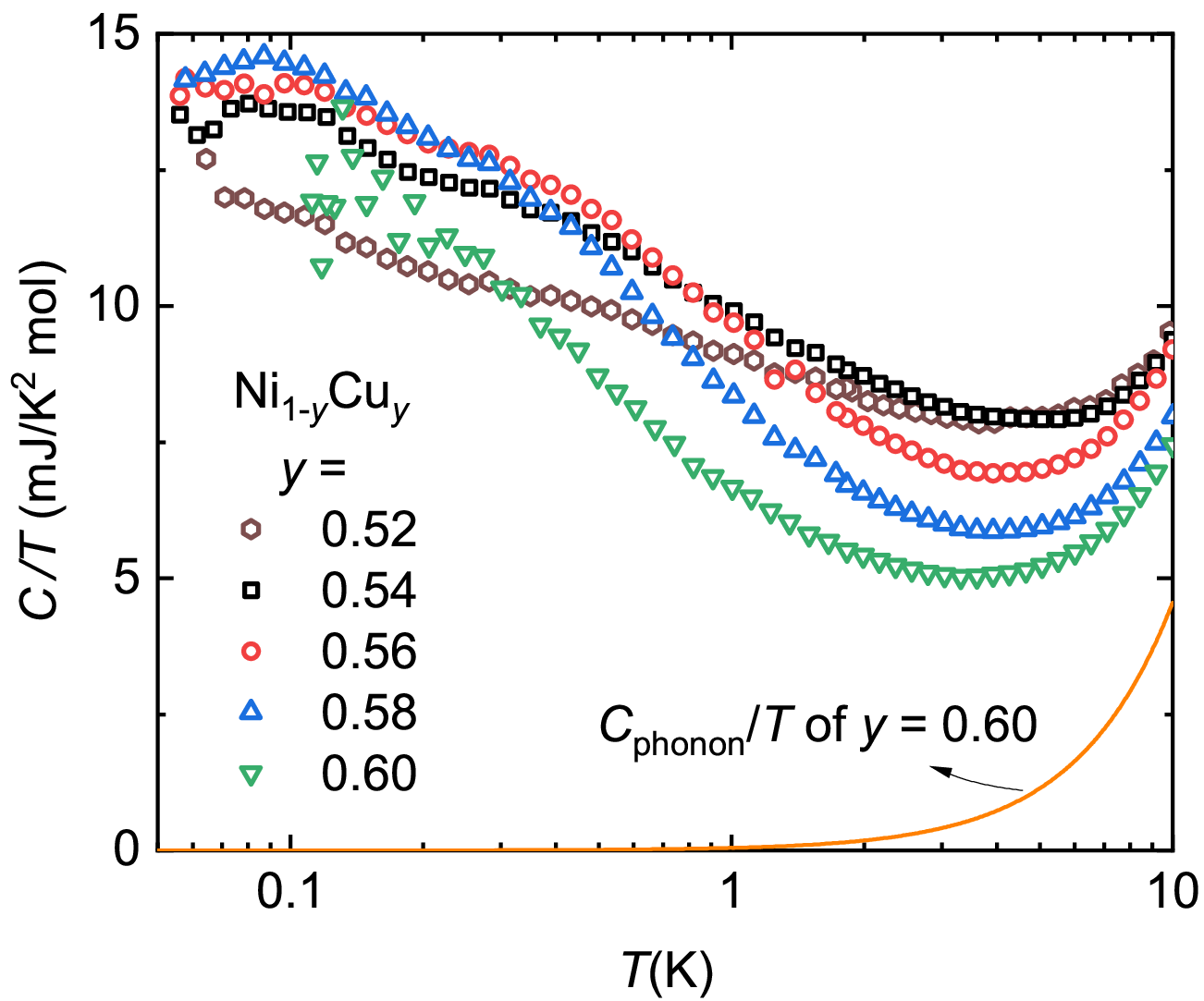}
\caption{Specific heat of Ni$_{1-y}$Cu$_{y}$ with $y = 0.52-0.60$ plotted as $C/T$ vs. log$T$. The solid line represents the phonon contribution to $C/T$ of the $y = 0.60$ sample, which is determined from high-temperature data with a Debye temperature of 342~K.}  
\label{CTDR} 
\end{figure}

For Ni$_{1-x}$Mo$_{x}$, it will be shown in the next section that the above mentioned quantum fluctuations are cut off by the appearance of spin glass behavior, and thus the FM QCP is avoided. For Ni$_{1-y}$Cu$_{y}$, we did not observe spin glass behavior down to 2~K. Zero-field $C/T$ measurements close to $y_{cr}$ are extended down to 0.06~K, as shown in Fig.~\ref{CTDR}. It is evident that all $C/T$ data diverge logarithmically in temperature down to 0.1 K at varying rates. Such divergence of $C/T$, which is irrelevant of phonon contributions, further strengthens the idea that the plausible FM QCP may exist in Ni$_{1-y}$Cu$_{y}$. This is akin to its sibling Ni$_{1-x}$Rh$_{x}$, where the $-$log$T$ divergence of $C/T$ was observed in the vicinity of the confirmed FM QCP \cite{Huang2020}. We observe that $C/T$ levels off below 0.1~K for $y = 0.54-0.58$. At the current stage, it is unclear whether this phenomenon is a result of resolution-limited measurements for small $C$ (although $C/T$ is larger at lower temperatures), or if it indicates that other phases terminate the effect of quantum fluctuations and hence the system is not in the critical region. Additional thermodynamic probes, such as thermal expansion, will be helpful in help resolve this issue.  

\subsection{Spin glass in Ni$_{1-x}$Mo$_{x}$} 
\begin{figure}
\includegraphics[width=\columnwidth]{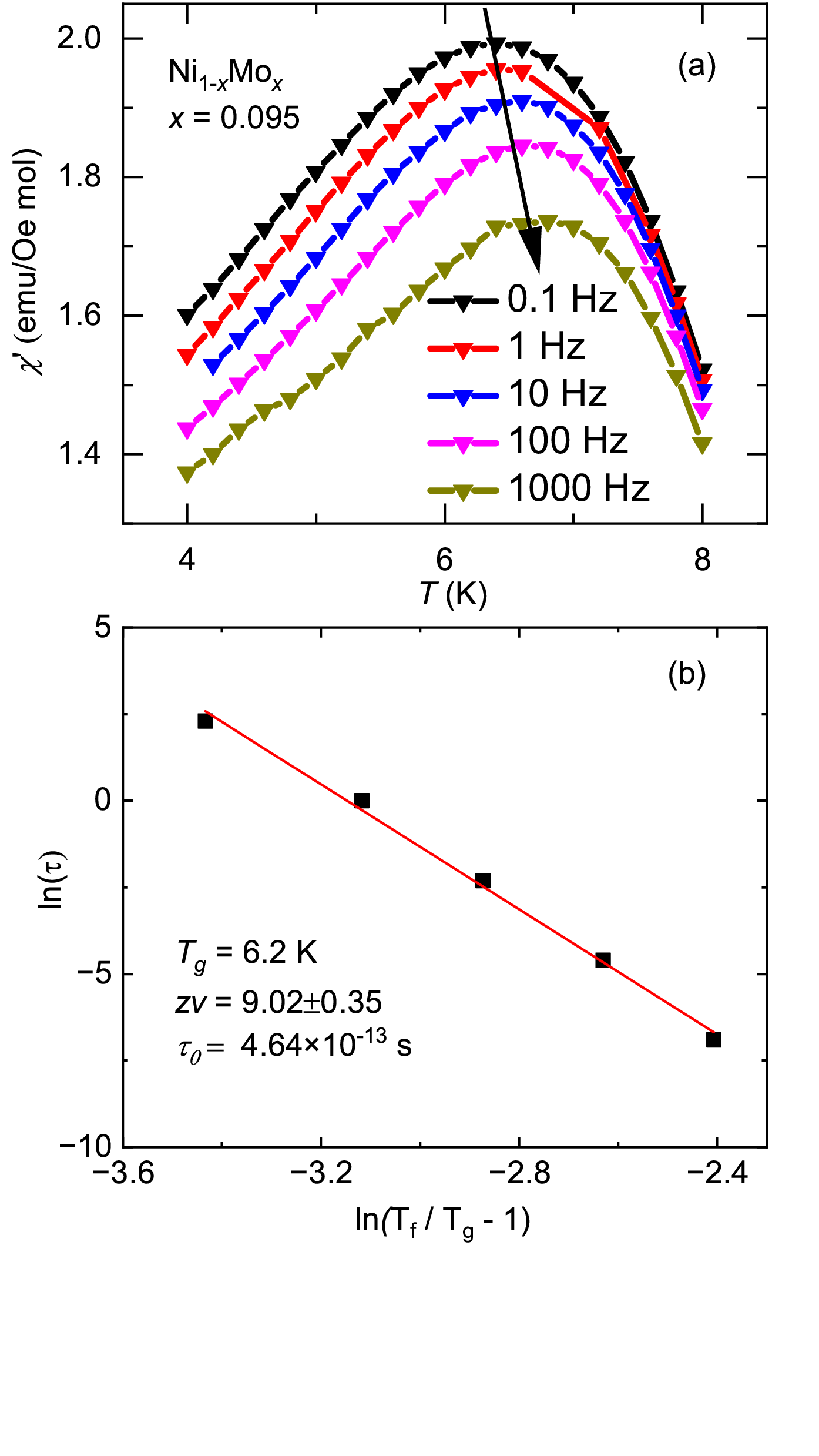}
\caption{(a) Temperature dependence of the real part ($\chi'$) of the ac magnetic susceptibility of Ni$_{1-x}$Mo$_{x}$ with $x = 0.095$. (b) Liner dynamical scaling of the relaxation time vs. the reduced temperature. The inset shows $R^{2}$ values obtained from linear fits in the main panel with different $T_{g}$'s.}  
\label{chiac} 
\end{figure}

It is known that the chemical replacement in a magnetic system can lead to strong disorder, subsequently causing spin glass behavior. This phenomenon has been observed in several ferromagnets as the chemical substitution is introduced to suppress the $T_{\rm C}$ \cite{Brando2016}. We have examined Ni$_{1-x}$Mo$_{x}$ and Ni$_{1-y}$Cu$_{y}$ down to 2~K. In Ni$_{1-y}$Cu$_{y}$ with $y = 0.54$, we did not observe the spin glass behavior, as shown in Fig.~\ref{chiacNC} in Appendix. For Ni$_{1-x}$Mo$_{x}$ with $x = 0.095$, we indeed found spin glass behavior via the ac magnetic susceptibility measurements. Figure~\ref{chiac}(a) shows the temperature dependence of the real part ($\chi'$) of the ac magnetic susceptibility where an ac field of 2.5~Oe with the frequency $f$ varies between 0.1 and 1000~Hz was applied. A peak in $\chi'(T)$ centered around $T_{f} = 6.3$~K for $f = 0.1$~Hz is observed. As $f$ increases, $T_{f}$ increases, and the overall intensity of $\chi'(T)$ decreases. This phenomenon is exactly how a spin glass system manifests in $\chi'(T)$ \cite{Mydosh1993}.  We can quantify the extent of peak shift by utilizing the Mydosh parameter ($S$) and categorize the class of the spin glass accordingly,
\begin{equation}
 S = \Delta T_{f}/T_{f}\cdot\Delta (\rm{log}_{10}\mathit{f}),
    \label{Mydosh}
\end{equation}
where $\Delta T_{f} = T_{f}(f_{1})-T_{f}(f_{2})$ and $\Delta (\rm{log}_{10}\mathit{f}) = \rm{log}_{10}\mathit{(f_{1})}-\rm{log}_{10}\mathit{(f_{2})}$ with $f_{1}$ = 1000~Hz and $f_{2}$ = 0.1~Hz. The calculated $S = 0.0133$ falls within the range of 0.005-0.08, and thus imply Ni$_{1-x}$Mo$_{x}$ with $x = 0.095$ belongs to the canonical spin-glass system \cite{Mydosh1993}. Our result agrees well with the theoretical calculations which have predicted the existence of spin glass behavior in Ni$_{1-x}$Mo$_{x}$ within the range of $x = 8-12$\% \cite{Ghosh1998}. 

To understand the dynamical properties of the spin glass phenomenon in Ni$_{1-x}$Mo$_{x}$, we applied the standard theory for dynamical scaling near $T_{\rm C}$. The conventional result of the dynamical scaling establishes a relationship between the critical relaxation time $\tau$ and the correlation length $\xi$ as $\tau \sim \xi^{z}$, and $\xi$ diverges with the temperature as $\xi \sim (T/(T-T_{\rm C}))^{\nu}$. In the context of a spin glass, the above equation can be expressed as:
\begin{equation}
  \tau = \tau_{0}(\frac{T_{f}}{T_{g}}-1)^{-z\nu},
    \label{Dyn1}
\end{equation}
where $\tau$ is the relaxation time with dependence on the measurement frequency ($\tau = 1/f$), $\tau_{0}$ is the characteristic relaxation time of a single spin flip, $T_{f}$ is the freezing temperature below which the moments freeze, $T_{g}$ is the freezing temperature as $f$ approaches zero, and $z\nu$ is the dynamical exponent. As indicated by Eq.~\ref{Dyn1}, a divergence occurs as $f$ decreases, resulting in multiple local minima that make the fitting challenging. Therefore, it is common to rewrite the Eq.~\ref{Dyn1} into a linear relationship \cite{Anand2012,Georg2022},
\begin{equation}
\textup{ln}(\tau)=\textup{ln}(\tau_{0})-z\nu\textup{ln}(\frac{T_{f}}{T_{g}}-1).
    \label{Dyn2}
\end{equation}
We chose $T_{g} = 6.2$~K just below $T_{f} = 6.3$~K. The obtained value for $\tau_{0}$ is $4.64\times 10^{-13}$ s, and the value of $z\nu$ is 9.02$\pm$0.35, as shown in Fig.~\ref{chiac}(b). For canonical spin glass systems, $\tau_{0}$ is typically between $10^{-12}$ and $10^{-14}$ s, while for cluster glass systems, $\tau_{0}$ falls within the range of $10^{-7}$ to $10^{-10}$ s \cite{Mydosh1993}. Our result indicates that Ni$_{1-x}$Mo$_{x}$ with $x = 0.095$ belongs to the canonical spin glass system. Besides, our $z\nu$ value falls with the range of 4-12 for most of spin glass systems \cite{Souletie1985}.     

\section{SUMMARY}
We have studied the magnetic properties of binary nickel alloys Ni$_{1-x}$Mo$_{x}$ and Ni$_{1-y}$Cu$_{y}$ via the structural analysis, electrical resistivity, magnetic susceptibility, critical exponents analysis based on isothermal magnetization curves, specific heat, and the ac magnetic susceptibility measurements. Our aim is to search for FM QCPs among binary nickel alloys. In Ni$_{1-x}$Mo$_{x}$, the FM QCP is prohibited due to the spin glass behavior near $x = x_{cr}$. The same outcome has been observed in Ni$_{1-x}$V$_{x}$ \cite{Ubaid-Kassis2010}, Ni$_{1-x}$Pd$_{x}$ \cite{Nicklas1999,Kuechler2006}, and other Ce- and U-based ferromagnets with composition tuning \cite{Westerkamp2009,Kawasaki2009,Pikul2012,Pikul20122}. 

For Ni$_{1-y}$Cu$_{y}$, after examining the $\rho(T)$ plot in Fig.\ref{rho}(b) and bearing the fact of the larger error bars in EPMA (Fig.\ref{EPMA}(b) in APPENDIX A), it seems that Ni$_{1-y}$Cu$_{y}$ exhibits a much stronger disorder effect and sample inhomogeneity compared to Ni$_{1-x}$Mo$_{x}$. However, spin glass or short-range order is not observed down to 1.8~K in Ni$_{1-y}$Cu$_{y}$. To address this concern, conducting ac magnetic susceptibility measurements at lower temperatures would provide further clarification. The logarithmic divergence of $C/T$ down to 0.1~K in the vicinity of $y_{cr}$ is observed and can be attributed to the effect of quantum fluctuations. However, for $T < 0.1$~K, the divergence of $C/T$ plateaus, preventing us from conclusively determining the existence of the FM QCP in Ni$_{1-y}$Cu$_{y}$. To shed light on this issue, other thermodynamic measurements at low temperatures are needed.

\section*{Acknowledgements}
We thank C.-C. Yang at SQUID VSM Lab, Instrumentation Center, National Cheng Kung University (NCKU), and Ms. J. Kang at Center for Condensed Matter Sciences (CCMS), National Taiwan University (NTU) for technical support. We also thank C.-W. Wang and C.-M. Wu at NSRRC for the preliminary neutron experiments at ANSTO. This work is supported by the National Science and Technology Council in Taiwan (grant numbers NSTC 109-2112-M-006-026-MY3, 110-2124-M-006-009) and the Higher Education Sprout Project, Ministry of Education to the Headquarters of University Advancement at NCKU. WTC acknowledges the National Science and Technology Council of Taiwan, for funding 111-2112-M-002-044-MY3, 112-2124-M-002-012, the Featured Areas Research Center Program within the framework of the Higher Education Sprout Project by the Ministry of Education of Taiwan 112L900802, and Academia Sinica project number AS-iMATE-111-12.

\section*{APPENDIX}
\subsection{Electron microscope microanalysis} 

\begin{figure}
\includegraphics[width=\columnwidth]{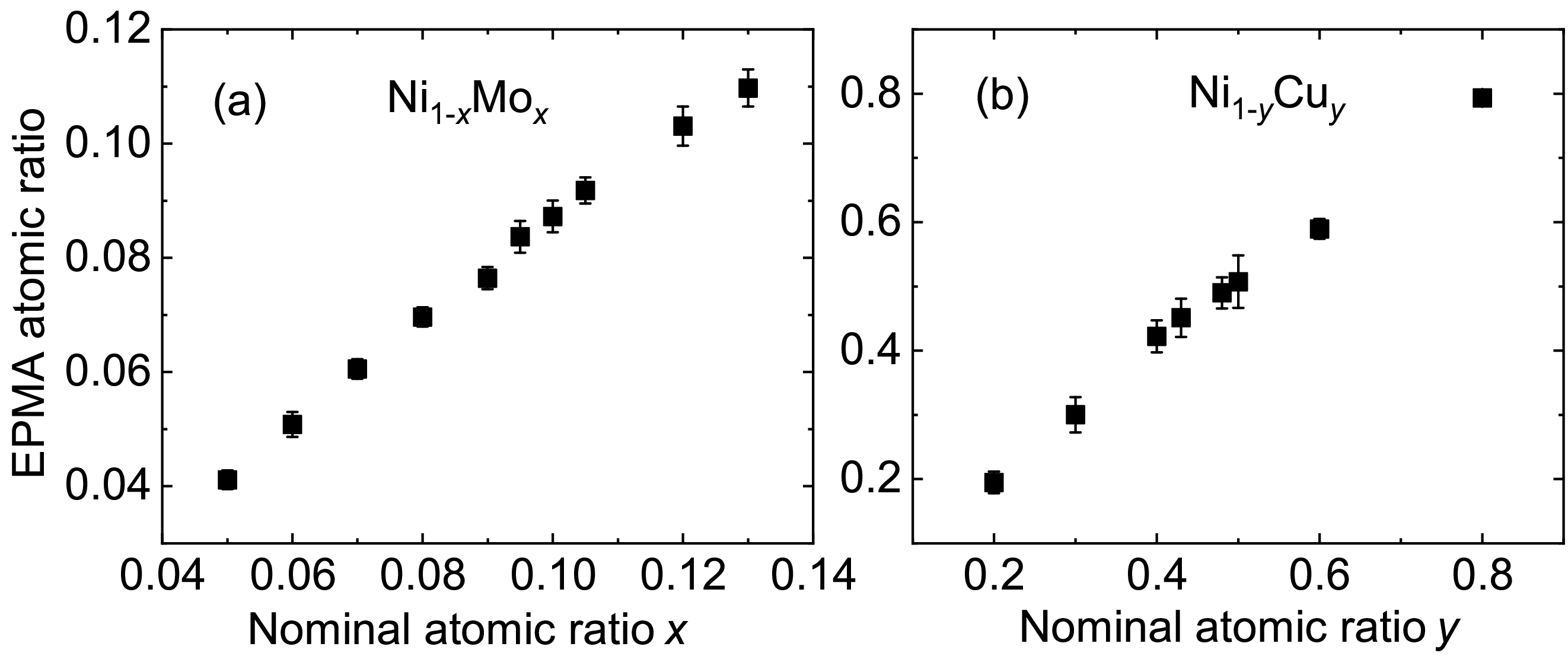}
\caption{(a) Mo concentrations in Ni$_{1-x}$Mo$_{x}$ and (b) Cu concentrations in Ni$_{1-y}$Cu$_{y}$ determined by EPMA vs.
nominal concentrations.}  
\label{EPMA} 
\end{figure}

For EPMA, as shown in Fig.~\ref{EPMA}, we measured 20 data points across a flat surface of each sample and the standard deviation in composition is $\leq$2\% for Ni$_{1-x}$Mo$_{x}$ and $\leq$10\% for Ni$_{1-y}$Cu$_{y}$. The largest standard deviation in Ni$_{1-y}$Cu$_{y}$ is found around $y = 0.50$, reflecting samples' inhomogeneity.

\subsection{$T_{\rm C}$ determined from the derivative of magnetic susceptibility}

We also use the temperatures at which the maximum negative slope in Fig.~\ref{chiT}(a) and (c) occurs as Tc , and the obtained values are consistent with the ones obtained using modified Arrott plots.

\subsection{ac magnetic susceptibility of Ni$_{1-y}$Cu$_{y}$}

\begin{figure}
\includegraphics[width=\columnwidth]{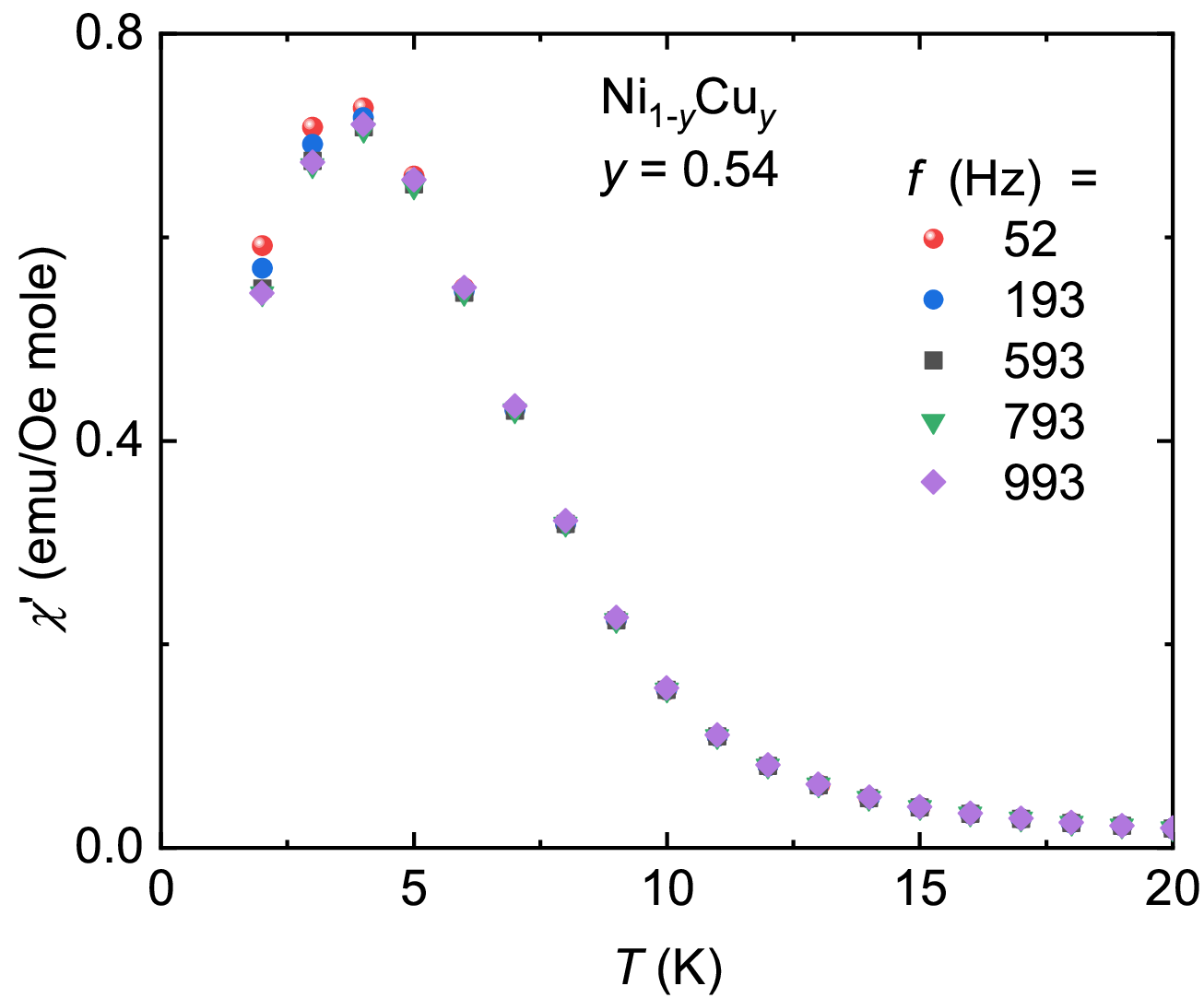}
\caption{Temperature dependence of the real part ($\chi'$) of the ac magnetic susceptibility of Ni$_{1-y}$Cu$_{y}$ with $y = 0.54$.}  
\label{chiacNC} 
\end{figure}

Figure~\ref{chiacNC} shows the ac magnetic susceptibility of Ni$_{1-y}$Cu$_{y}$ with $y = 0.54$. The low-temperature hump does not show frequency dependence that the hump remains at around 3.5~K for frequencies of 52-993 Hz. 

\begin{thebibliography}{43}%
\makeatletter
\providecommand \@ifxundefined [1]{%
 \@ifx{#1\undefined}
}%
\providecommand \@ifnum [1]{%
 \ifnum #1\expandafter \@firstoftwo
 \else \expandafter \@secondoftwo
 \fi
}%
\providecommand \@ifx [1]{%
 \ifx #1\expandafter \@firstoftwo
 \else \expandafter \@secondoftwo
 \fi
}%
\providecommand \natexlab [1]{#1}%
\providecommand \enquote  [1]{``#1''}%
\providecommand \bibnamefont  [1]{#1}%
\providecommand \bibfnamefont [1]{#1}%
\providecommand \citenamefont [1]{#1}%
\providecommand \href@noop [0]{\@secondoftwo}%
\providecommand \href [0]{\begingroup \@sanitize@url \@href}%
\providecommand \@href[1]{\@@startlink{#1}\@@href}%
\providecommand \@@href[1]{\endgroup#1\@@endlink}%
\providecommand \@sanitize@url [0]{\catcode `\\12\catcode `\$12\catcode `\&12\catcode `\#12\catcode `\^12\catcode `\_12\catcode `\%12\relax}%
\providecommand \@@startlink[1]{}%
\providecommand \@@endlink[0]{}%
\providecommand \url  [0]{\begingroup\@sanitize@url \@url }%
\providecommand \@url [1]{\endgroup\@href {#1}{\urlprefix }}%
\providecommand \urlprefix  [0]{URL }%
\providecommand \Eprint [0]{\href }%
\providecommand \doibase [0]{https://doi.org/}%
\providecommand \selectlanguage [0]{\@gobble}%
\providecommand \bibinfo  [0]{\@secondoftwo}%
\providecommand \bibfield  [0]{\@secondoftwo}%
\providecommand \translation [1]{[#1]}%
\providecommand \BibitemOpen [0]{}%
\providecommand \bibitemStop [0]{}%
\providecommand \bibitemNoStop [0]{.\EOS\space}%
\providecommand \EOS [0]{\spacefactor3000\relax}%
\providecommand \BibitemShut  [1]{\csname bibitem#1\endcsname}%
\let\auto@bib@innerbib\@empty
\bibitem [{\citenamefont {Brando}\ \emph {et~al.}(2016)\citenamefont {Brando}, \citenamefont {Belitz}, \citenamefont {Grosche},\ and\ \citenamefont {Kirkpatrick}}]{Brando2016}%
  \BibitemOpen
  \bibfield  {author} {\bibinfo {author} {\bibfnamefont {M.}~\bibnamefont {Brando}}, \bibinfo {author} {\bibfnamefont {D.}~\bibnamefont {Belitz}}, \bibinfo {author} {\bibfnamefont {F.~M.}\ \bibnamefont {Grosche}},\ and\ \bibinfo {author} {\bibfnamefont {T.~R.}\ \bibnamefont {Kirkpatrick}},\ }\bibfield  {title} {\bibinfo {title} {Metallic quantum ferromagnets},\ }\href {https://doi.org/10.1103/RevModPhys.88.025006} {\bibfield  {journal} {\bibinfo  {journal} {Rev. Mod. Phys.}\ }\textbf {\bibinfo {volume} {88}},\ \bibinfo {pages} {025006} (\bibinfo {year} {2016})}\BibitemShut {NoStop}%
\bibitem [{\citenamefont {Stewart}(2001)}]{Stewart2001}%
  \BibitemOpen
  \bibfield  {author} {\bibinfo {author} {\bibfnamefont {G.~R.}\ \bibnamefont {Stewart}},\ }\bibfield  {title} {\bibinfo {title} {Non-fermi-liquid behavior in $d$- and $f$-electron metals},\ }\href {https://doi.org/10.1103/RevModPhys.73.797} {\bibfield  {journal} {\bibinfo  {journal} {Rev. Mod. Phys.}\ }\textbf {\bibinfo {volume} {73}},\ \bibinfo {pages} {797} (\bibinfo {year} {2001})}\BibitemShut {NoStop}%
\bibitem [{\citenamefont {L{\"o}hneysen}\ \emph {et~al.}(2007)\citenamefont {L{\"o}hneysen}, \citenamefont {Rosch}, \citenamefont {Vojta},\ and\ \citenamefont {W{\"o}lfle}}]{Loehneysen2007}%
  \BibitemOpen
  \bibfield  {author} {\bibinfo {author} {\bibfnamefont {H.~v.}\ \bibnamefont {L{\"o}hneysen}}, \bibinfo {author} {\bibfnamefont {A.}~\bibnamefont {Rosch}}, \bibinfo {author} {\bibfnamefont {M.}~\bibnamefont {Vojta}},\ and\ \bibinfo {author} {\bibfnamefont {P.}~\bibnamefont {W{\"o}lfle}},\ }\bibfield  {title} {\bibinfo {title} {Fermi-liquid instabilities at magnetic quantum phase transitions},\ }\href@noop {} {\bibfield  {journal} {\bibinfo  {journal} {Reviews of Modern Physics}\ }\textbf {\bibinfo {volume} {79}},\ \bibinfo {pages} {1015} (\bibinfo {year} {2007})}\BibitemShut {NoStop}%
\bibitem [{\citenamefont {Huang}\ \emph {et~al.}(2016)\citenamefont {Huang}, \citenamefont {Eley}, \citenamefont {Rosa}, \citenamefont {Civale}, \citenamefont {Bauer}, \citenamefont {Baumbach}, \citenamefont {Maple},\ and\ \citenamefont {Janoschek}}]{Huang2016}%
  \BibitemOpen
  \bibfield  {author} {\bibinfo {author} {\bibfnamefont {K.}~\bibnamefont {Huang}}, \bibinfo {author} {\bibfnamefont {S.}~\bibnamefont {Eley}}, \bibinfo {author} {\bibfnamefont {P.~F.~S.}\ \bibnamefont {Rosa}}, \bibinfo {author} {\bibfnamefont {L.}~\bibnamefont {Civale}}, \bibinfo {author} {\bibfnamefont {E.~D.}\ \bibnamefont {Bauer}}, \bibinfo {author} {\bibfnamefont {R.~E.}\ \bibnamefont {Baumbach}}, \bibinfo {author} {\bibfnamefont {M.~B.}\ \bibnamefont {Maple}},\ and\ \bibinfo {author} {\bibfnamefont {M.}~\bibnamefont {Janoschek}},\ }\bibfield  {title} {\bibinfo {title} {Quantum critical scaling in the disordered itinerant ferromagnet ${\mathrm{uco}}_{1\ensuremath{-}x}{\mathrm{fe}}_{x}\mathrm{Ge}$},\ }\href {https://doi.org/10.1103/PhysRevLett.117.237202} {\bibfield  {journal} {\bibinfo  {journal} {Phys. Rev. Lett.}\ }\textbf {\bibinfo {volume} {117}},\ \bibinfo {pages} {237202} (\bibinfo {year} {2016})}\BibitemShut {NoStop}%
\bibitem [{\citenamefont {Goko}\ \emph {et~al.}(2017)\citenamefont {Goko}, \citenamefont {Arguello}, \citenamefont {Hamann}, \citenamefont {Wolf}, \citenamefont {Lee}, \citenamefont {Reznik}, \citenamefont {Maisuradze}, \citenamefont {Khasanov}, \citenamefont {Morenzoni},\ and\ \citenamefont {Uemura}}]{Goko2017}%
  \BibitemOpen
  \bibfield  {author} {\bibinfo {author} {\bibfnamefont {T.}~\bibnamefont {Goko}}, \bibinfo {author} {\bibfnamefont {C.~J.}\ \bibnamefont {Arguello}}, \bibinfo {author} {\bibfnamefont {A.}~\bibnamefont {Hamann}}, \bibinfo {author} {\bibfnamefont {T.}~\bibnamefont {Wolf}}, \bibinfo {author} {\bibfnamefont {M.}~\bibnamefont {Lee}}, \bibinfo {author} {\bibfnamefont {D.}~\bibnamefont {Reznik}}, \bibinfo {author} {\bibfnamefont {A.}~\bibnamefont {Maisuradze}}, \bibinfo {author} {\bibfnamefont {R.}~\bibnamefont {Khasanov}}, \bibinfo {author} {\bibfnamefont {E.}~\bibnamefont {Morenzoni}},\ and\ \bibinfo {author} {\bibfnamefont {Y.~J.}\ \bibnamefont {Uemura}},\ }\bibfield  {title} {\bibinfo {title} {Restoration of quantum critical behavior by disorder in pressure-tuned (mn,fe)si},\ }\href@noop {} {\bibfield  {journal} {\bibinfo  {journal} {npj Quantum Materials}\ }\textbf {\bibinfo {volume} {2}},\ \bibinfo {pages} {44} (\bibinfo {year} {2017})}\BibitemShut {NoStop}%
\bibitem [{\citenamefont {Sales}\ \emph {et~al.}(2017)\citenamefont {Sales}, \citenamefont {Jin}, \citenamefont {Bei}, \citenamefont {Nichols}, \citenamefont {Chisholm}, \citenamefont {May}, \citenamefont {Butch}, \citenamefont {Christianson},\ and\ \citenamefont {McGuire}}]{Sales2017}%
  \BibitemOpen
  \bibfield  {author} {\bibinfo {author} {\bibfnamefont {B.~C.}\ \bibnamefont {Sales}}, \bibinfo {author} {\bibfnamefont {K.}~\bibnamefont {Jin}}, \bibinfo {author} {\bibfnamefont {H.}~\bibnamefont {Bei}}, \bibinfo {author} {\bibfnamefont {J.}~\bibnamefont {Nichols}}, \bibinfo {author} {\bibfnamefont {M.~F.}\ \bibnamefont {Chisholm}}, \bibinfo {author} {\bibfnamefont {A.~F.}\ \bibnamefont {May}}, \bibinfo {author} {\bibfnamefont {N.~P.}\ \bibnamefont {Butch}}, \bibinfo {author} {\bibfnamefont {A.~D.}\ \bibnamefont {Christianson}},\ and\ \bibinfo {author} {\bibfnamefont {M.~A.}\ \bibnamefont {McGuire}},\ }\bibfield  {title} {\bibinfo {title} {Quantum critical behavior in the asymptotic limit of high disorder in the medium entropy alloy nicocr0.8},\ }\href@noop {} {\bibfield  {journal} {\bibinfo  {journal} {npj Quantum Materials}\ }\textbf {\bibinfo {volume} {2}},\ \bibinfo {pages} {33} (\bibinfo {year} {2017})}\BibitemShut {NoStop}%
\bibitem [{\citenamefont {Lai}\ \emph {et~al.}(2018)\citenamefont {Lai}, \citenamefont {Bone}, \citenamefont {Minasian}, \citenamefont {Ferrier}, \citenamefont {Lezama-Pacheco}, \citenamefont {Mocko}, \citenamefont {Ditter}, \citenamefont {Kozimor}, \citenamefont {Seidler}, \citenamefont {Nelson}, \citenamefont {Chiu}, \citenamefont {Huang}, \citenamefont {Potter}, \citenamefont {Graf}, \citenamefont {Albrecht-Schmitt},\ and\ \citenamefont {Baumbach}}]{Lai2018}%
  \BibitemOpen
  \bibfield  {author} {\bibinfo {author} {\bibfnamefont {Y.}~\bibnamefont {Lai}}, \bibinfo {author} {\bibfnamefont {S.~E.}\ \bibnamefont {Bone}}, \bibinfo {author} {\bibfnamefont {S.}~\bibnamefont {Minasian}}, \bibinfo {author} {\bibfnamefont {M.~G.}\ \bibnamefont {Ferrier}}, \bibinfo {author} {\bibfnamefont {J.}~\bibnamefont {Lezama-Pacheco}}, \bibinfo {author} {\bibfnamefont {V.}~\bibnamefont {Mocko}}, \bibinfo {author} {\bibfnamefont {A.~S.}\ \bibnamefont {Ditter}}, \bibinfo {author} {\bibfnamefont {S.~A.}\ \bibnamefont {Kozimor}}, \bibinfo {author} {\bibfnamefont {G.~T.}\ \bibnamefont {Seidler}}, \bibinfo {author} {\bibfnamefont {W.~L.}\ \bibnamefont {Nelson}}, \bibinfo {author} {\bibfnamefont {Y.-C.}\ \bibnamefont {Chiu}}, \bibinfo {author} {\bibfnamefont {K.}~\bibnamefont {Huang}}, \bibinfo {author} {\bibfnamefont {W.}~\bibnamefont {Potter}}, \bibinfo {author} {\bibfnamefont {D.}~\bibnamefont {Graf}}, \bibinfo {author} {\bibfnamefont {T.~E.}\ \bibnamefont {Albrecht-Schmitt}},\ and\ \bibinfo {author}
  {\bibfnamefont {R.~E.}\ \bibnamefont {Baumbach}},\ }\bibfield  {title} {\bibinfo {title} {Ferromagnetic quantum critical point in ${\mathrm{cepd}}_{2}{\mathrm{p}}_{2}$ with pd $\ensuremath{\rightarrow}$ ni substitution},\ }\href {https://doi.org/10.1103/PhysRevB.97.224406} {\bibfield  {journal} {\bibinfo  {journal} {Phys. Rev. B}\ }\textbf {\bibinfo {volume} {97}},\ \bibinfo {pages} {224406} (\bibinfo {year} {2018})}\BibitemShut {NoStop}%
\bibitem [{\citenamefont {Huang}\ \emph {et~al.}(2020)\citenamefont {Huang}, \citenamefont {Hallas}, \citenamefont {Grube}, \citenamefont {Kuntz}, \citenamefont {Spie\ss{}}, \citenamefont {Bayliff}, \citenamefont {Besara}, \citenamefont {Siegrist}, \citenamefont {Cai}, \citenamefont {Beare}, \citenamefont {Luke},\ and\ \citenamefont {Morosan}}]{Huang2020}%
  \BibitemOpen
  \bibfield  {author} {\bibinfo {author} {\bibfnamefont {C.-L.}\ \bibnamefont {Huang}}, \bibinfo {author} {\bibfnamefont {A.~M.}\ \bibnamefont {Hallas}}, \bibinfo {author} {\bibfnamefont {K.}~\bibnamefont {Grube}}, \bibinfo {author} {\bibfnamefont {S.}~\bibnamefont {Kuntz}}, \bibinfo {author} {\bibfnamefont {B.}~\bibnamefont {Spie\ss{}}}, \bibinfo {author} {\bibfnamefont {K.}~\bibnamefont {Bayliff}}, \bibinfo {author} {\bibfnamefont {T.}~\bibnamefont {Besara}}, \bibinfo {author} {\bibfnamefont {T.}~\bibnamefont {Siegrist}}, \bibinfo {author} {\bibfnamefont {Y.}~\bibnamefont {Cai}}, \bibinfo {author} {\bibfnamefont {J.}~\bibnamefont {Beare}}, \bibinfo {author} {\bibfnamefont {G.~M.}\ \bibnamefont {Luke}},\ and\ \bibinfo {author} {\bibfnamefont {E.}~\bibnamefont {Morosan}},\ }\bibfield  {title} {\bibinfo {title} {Quantum critical point in the itinerant ferromagnet ${\mathrm{ni}}_{1\ensuremath{-}x}{\mathrm{rh}}_{x}$},\ }\href@noop {} {\bibfield  {journal} {\bibinfo  {journal} {Phys. Rev. Lett.}\ }\textbf
  {\bibinfo {volume} {124}},\ \bibinfo {pages} {117203} (\bibinfo {year} {2020})}\BibitemShut {NoStop}%
\bibitem [{\citenamefont {Belitz}\ \emph {et~al.}(1999)\citenamefont {Belitz}, \citenamefont {Kirkpatrick},\ and\ \citenamefont {Vojta}}]{Belitz1999}%
  \BibitemOpen
  \bibfield  {author} {\bibinfo {author} {\bibfnamefont {D.}~\bibnamefont {Belitz}}, \bibinfo {author} {\bibfnamefont {T.~R.}\ \bibnamefont {Kirkpatrick}},\ and\ \bibinfo {author} {\bibfnamefont {T.}~\bibnamefont {Vojta}},\ }\bibfield  {title} {\bibinfo {title} {First order transitions and multicritical points in weak itinerant ferromagnets},\ }\href {https://doi.org/10.1103/PhysRevLett.82.4707} {\bibfield  {journal} {\bibinfo  {journal} {Phys. Rev. Lett.}\ }\textbf {\bibinfo {volume} {82}},\ \bibinfo {pages} {4707} (\bibinfo {year} {1999})}\BibitemShut {NoStop}%
\bibitem [{\citenamefont {Sang}\ \emph {et~al.}(2014)\citenamefont {Sang}, \citenamefont {Belitz},\ and\ \citenamefont {Kirkpatrick}}]{Sang2014}%
  \BibitemOpen
  \bibfield  {author} {\bibinfo {author} {\bibfnamefont {Y.}~\bibnamefont {Sang}}, \bibinfo {author} {\bibfnamefont {D.}~\bibnamefont {Belitz}},\ and\ \bibinfo {author} {\bibfnamefont {T.~R.}\ \bibnamefont {Kirkpatrick}},\ }\bibfield  {title} {\bibinfo {title} {Disorder dependence of the ferromagnetic quantum phase transition},\ }\href {https://doi.org/10.1103/PhysRevLett.113.207201} {\bibfield  {journal} {\bibinfo  {journal} {Phys. Rev. Lett.}\ }\textbf {\bibinfo {volume} {113}},\ \bibinfo {pages} {207201} (\bibinfo {year} {2014})}\BibitemShut {NoStop}%
\bibitem [{\citenamefont {Kirkpatrick}\ and\ \citenamefont {Belitz}(2014)}]{Kirkpatrick2014}%
  \BibitemOpen
  \bibfield  {author} {\bibinfo {author} {\bibfnamefont {T.~R.}\ \bibnamefont {Kirkpatrick}}\ and\ \bibinfo {author} {\bibfnamefont {D.}~\bibnamefont {Belitz}},\ }\bibfield  {title} {\bibinfo {title} {Preasymptotic critical behavior and effective exponents in disordered metallic quantum ferromagnets},\ }\href {https://doi.org/10.1103/PhysRevLett.113.127203} {\bibfield  {journal} {\bibinfo  {journal} {Phys. Rev. Lett.}\ }\textbf {\bibinfo {volume} {113}},\ \bibinfo {pages} {127203} (\bibinfo {year} {2014})}\BibitemShut {NoStop}%
\bibitem [{\citenamefont {Kirkpatrick}\ and\ \citenamefont {Belitz}(2015)}]{Kirkpatrick2015}%
  \BibitemOpen
  \bibfield  {author} {\bibinfo {author} {\bibfnamefont {T.~R.}\ \bibnamefont {Kirkpatrick}}\ and\ \bibinfo {author} {\bibfnamefont {D.}~\bibnamefont {Belitz}},\ }\bibfield  {title} {\bibinfo {title} {Exponent relations at quantum phase transitions with applications to metallic quantum ferromagnets},\ }\href {https://doi.org/10.1103/PhysRevB.91.214407} {\bibfield  {journal} {\bibinfo  {journal} {Phys. Rev. B}\ }\textbf {\bibinfo {volume} {91}},\ \bibinfo {pages} {214407} (\bibinfo {year} {2015})}\BibitemShut {NoStop}%
\bibitem [{\citenamefont {Gupta}\ \emph {et~al.}(1964)\citenamefont {Gupta}, \citenamefont {Cheng},\ and\ \citenamefont {Beck}}]{Gupta1964}%
  \BibitemOpen
  \bibfield  {author} {\bibinfo {author} {\bibfnamefont {K.~P.}\ \bibnamefont {Gupta}}, \bibinfo {author} {\bibfnamefont {C.~H.}\ \bibnamefont {Cheng}},\ and\ \bibinfo {author} {\bibfnamefont {P.~A.}\ \bibnamefont {Beck}},\ }\bibfield  {title} {\bibinfo {title} {Low-temperature specific heat of {Ni}-base fcc solid solutions with {Cu}, {Zn}, {Al}, {Si}, and {Sb}},\ }\href {https://doi.org/10.1103/PhysRev.133.A203} {\bibfield  {journal} {\bibinfo  {journal} {Phys. Rev.}\ }\textbf {\bibinfo {volume} {133}},\ \bibinfo {pages} {A203} (\bibinfo {year} {1964})}\BibitemShut {NoStop}%
\bibitem [{\citenamefont {Brinkman}\ \emph {et~al.}(1968)\citenamefont {Brinkman}, \citenamefont {Bucher}, \citenamefont {Williams},\ and\ \citenamefont {Maita}}]{Brinkman1968}%
  \BibitemOpen
  \bibfield  {author} {\bibinfo {author} {\bibfnamefont {W.~F.}\ \bibnamefont {Brinkman}}, \bibinfo {author} {\bibfnamefont {E.}~\bibnamefont {Bucher}}, \bibinfo {author} {\bibfnamefont {H.~J.}\ \bibnamefont {Williams}},\ and\ \bibinfo {author} {\bibfnamefont {J.~P.}\ \bibnamefont {Maita}},\ }\bibfield  {title} {\bibinfo {title} {The magnetic susceptibility and specific heat of nearly ferromagnetic nirh alloys},\ }\href {https://doi.org/10.1063/1.2163512} {\bibfield  {journal} {\bibinfo  {journal} {Journal of Applied Physics}\ }\textbf {\bibinfo {volume} {39}},\ \bibinfo {pages} {547} (\bibinfo {year} {1968})}\BibitemShut {NoStop}%
\bibitem [{\citenamefont {Bo\"lling}(1968)}]{Boelling1968}%
  \BibitemOpen
  \bibfield  {author} {\bibinfo {author} {\bibfnamefont {F.}~\bibnamefont {Bo\"lling}},\ }\bibfield  {title} {\bibinfo {title} {Ferro- und paramagnetisches verhalten in den mischkristallen des nickels mit vanadium, rhodium und platin im temperaturbereich von {14$^{\circ}$} bis 1000{1000$^{\circ}$}k},\ }\href@noop {} {\bibfield  {journal} {\bibinfo  {journal} {Phys kondens Materie}\ }\textbf {\bibinfo {volume} {7}},\ \bibinfo {pages} {162} (\bibinfo {year} {1968})}\BibitemShut {NoStop}%
\bibitem [{\citenamefont {Gregory}\ and\ \citenamefont {Moody}(1975)}]{Gregory1975}%
  \BibitemOpen
  \bibfield  {author} {\bibinfo {author} {\bibfnamefont {I.~P.}\ \bibnamefont {Gregory}}\ and\ \bibinfo {author} {\bibfnamefont {D.~E.}\ \bibnamefont {Moody}},\ }\bibfield  {title} {\bibinfo {title} {The low temperature specific heat and magnetization of binary alloys of nickel with titanium, vanadium, chromium and manganese},\ }\href {https://doi.org/10.1088/0305-4608/5/1/008} {\bibfield  {journal} {\bibinfo  {journal} {Journal of Physics F: Metal Physics}\ }\textbf {\bibinfo {volume} {5}},\ \bibinfo {pages} {36} (\bibinfo {year} {1975})}\BibitemShut {NoStop}%
\bibitem [{\citenamefont {Nicklas}\ \emph {et~al.}(1999)\citenamefont {Nicklas}, \citenamefont {Brando}, \citenamefont {Knebel}, \citenamefont {Mayr}, \citenamefont {Trinkl},\ and\ \citenamefont {Loidl}}]{Nicklas1999}%
  \BibitemOpen
  \bibfield  {author} {\bibinfo {author} {\bibfnamefont {M.}~\bibnamefont {Nicklas}}, \bibinfo {author} {\bibfnamefont {M.}~\bibnamefont {Brando}}, \bibinfo {author} {\bibfnamefont {G.}~\bibnamefont {Knebel}}, \bibinfo {author} {\bibfnamefont {F.}~\bibnamefont {Mayr}}, \bibinfo {author} {\bibfnamefont {W.}~\bibnamefont {Trinkl}},\ and\ \bibinfo {author} {\bibfnamefont {A.}~\bibnamefont {Loidl}},\ }\bibfield  {title} {\bibinfo {title} {Non-{Fermi}-liquid behavior at a ferromagnetic quantum critical point in {Ni$_{1-x}$Pd$_x$}},\ }\href@noop {} {\bibfield  {journal} {\bibinfo  {journal} {Phys. Rev. Lett.}\ }\textbf {\bibinfo {volume} {82}},\ \bibinfo {pages} {4268} (\bibinfo {year} {1999})}\BibitemShut {NoStop}%
\bibitem [{\citenamefont {Ubaid-Kassis}\ \emph {et~al.}(2010)\citenamefont {Ubaid-Kassis}, \citenamefont {Vojta},\ and\ \citenamefont {Schroeder}}]{Ubaid-Kassis2010}%
  \BibitemOpen
  \bibfield  {author} {\bibinfo {author} {\bibfnamefont {S.}~\bibnamefont {Ubaid-Kassis}}, \bibinfo {author} {\bibfnamefont {T.}~\bibnamefont {Vojta}},\ and\ \bibinfo {author} {\bibfnamefont {A.}~\bibnamefont {Schroeder}},\ }\bibfield  {title} {\bibinfo {title} {Quantum griffiths phase in the weak itinerant ferromagnetic alloy {Ni$_{1-x}$V$_x$}},\ }\href@noop {} {\bibfield  {journal} {\bibinfo  {journal} {Phys. Rev. Lett.}\ }\textbf {\bibinfo {volume} {104}},\ \bibinfo {pages} {066402} (\bibinfo {year} {2010})}\BibitemShut {NoStop}%
\bibitem [{\citenamefont {Lin}\ and\ \citenamefont {Huang}(2022)}]{Lin2022}%
  \BibitemOpen
  \bibfield  {author} {\bibinfo {author} {\bibfnamefont {R.-Z.}\ \bibnamefont {Lin}}\ and\ \bibinfo {author} {\bibfnamefont {C.-L.}\ \bibnamefont {Huang}},\ }\bibfield  {title} {\bibinfo {title} {Hyperscaling analysis of a disorder-induced ferromagnetic quantum critical point in ${\mathrm{ni}}_{1\ensuremath{-}x}{\mathrm{rh}}_{x}$ with $x=0.375$},\ }\href {https://doi.org/10.1103/PhysRevB.105.024429} {\bibfield  {journal} {\bibinfo  {journal} {Phys. Rev. B}\ }\textbf {\bibinfo {volume} {105}},\ \bibinfo {pages} {024429} (\bibinfo {year} {2022})}\BibitemShut {NoStop}%
\bibitem [{\citenamefont {Turchanin}\ \emph {et~al.}(2007)\citenamefont {Turchanin}, \citenamefont {Agraval},\ and\ \citenamefont {Abdulov}}]{Turchanin2007}%
  \BibitemOpen
  \bibfield  {author} {\bibinfo {author} {\bibfnamefont {M.~A.}\ \bibnamefont {Turchanin}}, \bibinfo {author} {\bibfnamefont {P.~G.}\ \bibnamefont {Agraval}},\ and\ \bibinfo {author} {\bibfnamefont {A.~R.}\ \bibnamefont {Abdulov}},\ }\bibfield  {title} {\bibinfo {title} {Phase equilibria and thermodynamics of binary copper systems with 3d-metals. vi. copper-nickel system},\ }\href@noop {} {\bibfield  {journal} {\bibinfo  {journal} {Powder Metallurgy and Metal Ceramics}\ }\textbf {\bibinfo {volume} {46}},\ \bibinfo {pages} {467} (\bibinfo {year} {2007})}\BibitemShut {NoStop}%
\bibitem [{\citenamefont {D{\'i}az}\ \emph {et~al.}(2021)\citenamefont {D{\'i}az}, \citenamefont {Herrera}, \citenamefont {Oyarz{\'u}n},\ and\ \citenamefont {Munoz}}]{Diaz2021}%
  \BibitemOpen
  \bibfield  {author} {\bibinfo {author} {\bibfnamefont {E.}~\bibnamefont {D{\'i}az}}, \bibinfo {author} {\bibfnamefont {G.}~\bibnamefont {Herrera}}, \bibinfo {author} {\bibfnamefont {S.}~\bibnamefont {Oyarz{\'u}n}},\ and\ \bibinfo {author} {\bibfnamefont {R.~C.}\ \bibnamefont {Munoz}},\ }\bibfield  {title} {\bibinfo {title} {Evidence of weak anderson localization revealed by the resistivity, transverse magnetoresistance and hall effect measured on thin cu films deposited on mica},\ }\href@noop {} {\bibfield  {journal} {\bibinfo  {journal} {Scientific Reports}\ }\textbf {\bibinfo {volume} {11}},\ \bibinfo {pages} {17820} (\bibinfo {year} {2021})}\BibitemShut {NoStop}%
\bibitem [{\citenamefont {Amamou}\ \emph {et~al.}(1975)\citenamefont {Amamou}, \citenamefont {Gautier},\ and\ \citenamefont {B.}}]{Amamou1975}%
  \BibitemOpen
  \bibfield  {author} {\bibinfo {author} {\bibfnamefont {A.}~\bibnamefont {Amamou}}, \bibinfo {author} {\bibfnamefont {F.}~\bibnamefont {Gautier}},\ and\ \bibinfo {author} {\bibfnamefont {L.}~\bibnamefont {B.}},\ }\href {https://doi.org/10.1088/0305-4608/5/7/014} {\bibfield  {journal} {\bibinfo  {journal} {Journal of Physics F: Metal Physics}\ }\textbf {\bibinfo {volume} {5}},\ \bibinfo {pages} {1342} (\bibinfo {year} {1975})}\BibitemShut {NoStop}%
\bibitem [{\citenamefont {Santiago}\ \emph {et~al.}(2017)\citenamefont {Santiago}, \citenamefont {Huang},\ and\ \citenamefont {Morosan}}]{Santiago2017}%
  \BibitemOpen
  \bibfield  {author} {\bibinfo {author} {\bibfnamefont {J.~M.}\ \bibnamefont {Santiago}}, \bibinfo {author} {\bibfnamefont {C.-L.}\ \bibnamefont {Huang}},\ and\ \bibinfo {author} {\bibfnamefont {E.}~\bibnamefont {Morosan}},\ }\bibfield  {title} {\bibinfo {title} {Itinerant magnetic metals},\ }\href {https://doi.org/10.1088/1361-648X/aa7889} {\bibfield  {journal} {\bibinfo  {journal} {Journal of Physics: Condensed Matter}\ }\textbf {\bibinfo {volume} {29}},\ \bibinfo {pages} {373002} (\bibinfo {year} {2017})}\BibitemShut {NoStop}%
\bibitem [{\citenamefont {Mugiraneza}\ and\ \citenamefont {Hallas}(2022)}]{Mugiraneza2022}%
  \BibitemOpen
  \bibfield  {author} {\bibinfo {author} {\bibfnamefont {S.}~\bibnamefont {Mugiraneza}}\ and\ \bibinfo {author} {\bibfnamefont {A.~M.}\ \bibnamefont {Hallas}},\ }\bibfield  {title} {\bibinfo {title} {Tutorial: a beginner's guide to interpreting magnetic susceptibility data with the curie-weiss law},\ }\href@noop {} {\bibfield  {journal} {\bibinfo  {journal} {Communications Physics}\ }\textbf {\bibinfo {volume} {5}},\ \bibinfo {pages} {95} (\bibinfo {year} {2022})}\BibitemShut {NoStop}%
\bibitem [{\citenamefont {Fuchs}\ \emph {et~al.}(2014)\citenamefont {Fuchs}, \citenamefont {Wissinger}, \citenamefont {Schmalian}, \citenamefont {Huang}, \citenamefont {Fromknecht}, \citenamefont {Schneider},\ and\ \citenamefont {L\"ohneysen}}]{Fuchs2014}%
  \BibitemOpen
  \bibfield  {author} {\bibinfo {author} {\bibfnamefont {D.}~\bibnamefont {Fuchs}}, \bibinfo {author} {\bibfnamefont {M.}~\bibnamefont {Wissinger}}, \bibinfo {author} {\bibfnamefont {J.}~\bibnamefont {Schmalian}}, \bibinfo {author} {\bibfnamefont {C.-L.}\ \bibnamefont {Huang}}, \bibinfo {author} {\bibfnamefont {R.}~\bibnamefont {Fromknecht}}, \bibinfo {author} {\bibfnamefont {R.}~\bibnamefont {Schneider}},\ and\ \bibinfo {author} {\bibfnamefont {H.~v.}\ \bibnamefont {L\"ohneysen}},\ }\bibfield  {title} {\bibinfo {title} {Critical scaling analysis of the itinerant ferromagnet ${\mathrm{sr}}_{1\ensuremath{-}x}$${\mathrm{ca}}_{x}$${\mathrm{ruo}}_{3}$},\ }\href@noop {} {\bibfield  {journal} {\bibinfo  {journal} {Phys. Rev. B}\ }\textbf {\bibinfo {volume} {89}},\ \bibinfo {pages} {174405} (\bibinfo {year} {2014})}\BibitemShut {NoStop}%
\bibitem [{\citenamefont {Huang}\ \emph {et~al.}(2015)\citenamefont {Huang}, \citenamefont {Fuchs}, \citenamefont {Wissinger}, \citenamefont {Schneider}, \citenamefont {Ling}, \citenamefont {Scheurer}, \citenamefont {Schmalian},\ and\ \citenamefont {L{\"o}hneysen}}]{Huang2015}%
  \BibitemOpen
  \bibfield  {author} {\bibinfo {author} {\bibfnamefont {C.~L.}\ \bibnamefont {Huang}}, \bibinfo {author} {\bibfnamefont {D.}~\bibnamefont {Fuchs}}, \bibinfo {author} {\bibfnamefont {M.}~\bibnamefont {Wissinger}}, \bibinfo {author} {\bibfnamefont {R.}~\bibnamefont {Schneider}}, \bibinfo {author} {\bibfnamefont {M.~C.}\ \bibnamefont {Ling}}, \bibinfo {author} {\bibfnamefont {M.~S.}\ \bibnamefont {Scheurer}}, \bibinfo {author} {\bibfnamefont {J.}~\bibnamefont {Schmalian}},\ and\ \bibinfo {author} {\bibfnamefont {H.~v.}\ \bibnamefont {L{\"o}hneysen}},\ }\bibfield  {title} {\bibinfo {title} {Anomalous quantum criticality in an itinerant ferromagnet},\ }\href@noop {} {\bibfield  {journal} {\bibinfo  {journal} {Nature Communications}\ }\textbf {\bibinfo {volume} {6}},\ \bibinfo {pages} {8188} (\bibinfo {year} {2015})}\BibitemShut {NoStop}%
\bibitem [{\citenamefont {Sur}\ \emph {et~al.}(2023)\citenamefont {Sur}, \citenamefont {Kim}, \citenamefont {Kim},\ and\ \citenamefont {Kim}}]{Sur2023}%
  \BibitemOpen
  \bibfield  {author} {\bibinfo {author} {\bibfnamefont {Y.}~\bibnamefont {Sur}}, \bibinfo {author} {\bibfnamefont {K.-T.}\ \bibnamefont {Kim}}, \bibinfo {author} {\bibfnamefont {S.}~\bibnamefont {Kim}},\ and\ \bibinfo {author} {\bibfnamefont {K.~H.}\ \bibnamefont {Kim}},\ }\bibfield  {title} {\bibinfo {title} {Optimized superconductivity in the vicinity of a nematic quantum critical point in the kagome superconductor cs(v1-xtix)3sb5},\ }\href@noop {} {\bibfield  {journal} {\bibinfo  {journal} {Nature Communications}\ }\textbf {\bibinfo {volume} {14}},\ \bibinfo {pages} {3899} (\bibinfo {year} {2023})}\BibitemShut {NoStop}%
\bibitem [{\citenamefont {Arrott}\ and\ \citenamefont {Noakes}(1967)}]{Arrott1967}%
  \BibitemOpen
  \bibfield  {author} {\bibinfo {author} {\bibfnamefont {A.}~\bibnamefont {Arrott}}\ and\ \bibinfo {author} {\bibfnamefont {J.~E.}\ \bibnamefont {Noakes}},\ }\bibfield  {title} {\bibinfo {title} {Approximate equation of state for nickel near its critical temperature},\ }\href@noop {} {\bibfield  {journal} {\bibinfo  {journal} {Phys. Rev. Lett.}\ }\textbf {\bibinfo {volume} {19}},\ \bibinfo {pages} {786} (\bibinfo {year} {1967})}\BibitemShut {NoStop}%
\bibitem [{\citenamefont {Kaul}(1985)}]{Kaul1985}%
  \BibitemOpen
  \bibfield  {author} {\bibinfo {author} {\bibfnamefont {S.~N.}\ \bibnamefont {Kaul}},\ }\bibfield  {title} {\bibinfo {title} {Static critical phenomena in ferromagnets with quenched disorder},\ }\href@noop {} {\bibfield  {journal} {\bibinfo  {journal} {Journal of Magnetism and Magnetic Materials}\ }\textbf {\bibinfo {volume} {53}},\ \bibinfo {pages} {5} (\bibinfo {year} {1985})}\BibitemShut {NoStop}%
\bibitem [{\citenamefont {Stanley}(1971)}]{Stanley1971}%
  \BibitemOpen
  \bibfield  {author} {\bibinfo {author} {\bibfnamefont {H.~E.}\ \bibnamefont {Stanley}},\ }\href@noop {} {\emph {\bibinfo {title} {Introduction to phase transitions and critical phenomena}}}\ (\bibinfo  {publisher} {Oxford University Press, New York},\ \bibinfo {year} {1971})\BibitemShut {NoStop}%
\bibitem [{\citenamefont {St\"usser}\ \emph {et~al.}(1986)\citenamefont {St\"usser}, \citenamefont {Rekveldt},\ and\ \citenamefont {Spruijt}}]{Stuesser1986}%
  \BibitemOpen
  \bibfield  {author} {\bibinfo {author} {\bibfnamefont {N.}~\bibnamefont {St\"usser}}, \bibinfo {author} {\bibfnamefont {M.~T.}\ \bibnamefont {Rekveldt}},\ and\ \bibinfo {author} {\bibfnamefont {T.}~\bibnamefont {Spruijt}},\ }\bibfield  {title} {\bibinfo {title} {Critical exponents, amplitudes, and correction to scaling in nickel measured by neutron depolarization},\ }\href@noop {} {\bibfield  {journal} {\bibinfo  {journal} {Phys. Rev. B}\ }\textbf {\bibinfo {volume} {33}},\ \bibinfo {pages} {6423} (\bibinfo {year} {1986})}\BibitemShut {NoStop}%
\bibitem [{\citenamefont {Collins}(1989)}]{Collins1989}%
  \BibitemOpen
  \bibfield  {author} {\bibinfo {author} {\bibfnamefont {M.~F.}\ \bibnamefont {Collins}},\ }\href@noop {} {\emph {\bibinfo {title} {Magnetic Critical Scattering}}}\ (\bibinfo  {publisher} {Oxford University Press, Oxford Series on Neutron Scattering in Condensed Matter},\ \bibinfo {year} {1989})\BibitemShut {NoStop}%
\bibitem [{\citenamefont {Huy}\ \emph {et~al.}(2007)\citenamefont {Huy}, \citenamefont {Gasparini}, \citenamefont {Klaasse}, \citenamefont {de~Visser}, \citenamefont {Sakarya},\ and\ \citenamefont {van Dijk}}]{Huy2007}%
  \BibitemOpen
  \bibfield  {author} {\bibinfo {author} {\bibfnamefont {N.~T.}\ \bibnamefont {Huy}}, \bibinfo {author} {\bibfnamefont {A.}~\bibnamefont {Gasparini}}, \bibinfo {author} {\bibfnamefont {J.~C.~P.}\ \bibnamefont {Klaasse}}, \bibinfo {author} {\bibfnamefont {A.}~\bibnamefont {de~Visser}}, \bibinfo {author} {\bibfnamefont {S.}~\bibnamefont {Sakarya}},\ and\ \bibinfo {author} {\bibfnamefont {N.~H.}\ \bibnamefont {van Dijk}},\ }\bibfield  {title} {\bibinfo {title} {Ferromagnetic quantum critical point in urhge doped with ru},\ }\href {https://doi.org/10.1103/PhysRevB.75.212405} {\bibfield  {journal} {\bibinfo  {journal} {Phys. Rev. B}\ }\textbf {\bibinfo {volume} {75}},\ \bibinfo {pages} {212405} (\bibinfo {year} {2007})}\BibitemShut {NoStop}%
\bibitem [{\citenamefont {Mydosh}(1993)}]{Mydosh1993}%
  \BibitemOpen
  \bibfield  {author} {\bibinfo {author} {\bibfnamefont {J.~Y.}\ \bibnamefont {Mydosh}},\ }\href@noop {} {\emph {\bibinfo {title} {Spin Glasses: An Experimental Introduction (1st ed.).}}}\ (\bibinfo  {publisher} {CRC Press},\ \bibinfo {year} {1993})\BibitemShut {NoStop}%
\bibitem [{\citenamefont {Ghosh}\ \emph {et~al.}(1998)\citenamefont {Ghosh}, \citenamefont {Das},\ and\ \citenamefont {Mookerjee}}]{Ghosh1998}%
  \BibitemOpen
  \bibfield  {author} {\bibinfo {author} {\bibfnamefont {S.}~\bibnamefont {Ghosh}}, \bibinfo {author} {\bibfnamefont {N.}~\bibnamefont {Das}},\ and\ \bibinfo {author} {\bibfnamefont {A.}~\bibnamefont {Mookerjee}},\ }\bibfield  {title} {\bibinfo {title} {Magnetic properties of ni-mo single-crystal alloys; theory and experiment},\ }\href {https://doi.org/10.1088/0953-8984/10/50/016} {\bibfield  {journal} {\bibinfo  {journal} {Journal of Physics: Condensed Matter}\ }\textbf {\bibinfo {volume} {10}},\ \bibinfo {pages} {11773} (\bibinfo {year} {1998})}\BibitemShut {NoStop}%
\bibitem [{\citenamefont {Anand}\ \emph {et~al.}(2012)\citenamefont {Anand}, \citenamefont {Adroja},\ and\ \citenamefont {Hillier}}]{Anand2012}%
  \BibitemOpen
  \bibfield  {author} {\bibinfo {author} {\bibfnamefont {V.~K.}\ \bibnamefont {Anand}}, \bibinfo {author} {\bibfnamefont {D.~T.}\ \bibnamefont {Adroja}},\ and\ \bibinfo {author} {\bibfnamefont {A.~D.}\ \bibnamefont {Hillier}},\ }\bibfield  {title} {\bibinfo {title} {Ferromagnetic cluster spin-glass behavior in prrhsn${}_{3}$},\ }\href {https://doi.org/10.1103/PhysRevB.85.014418} {\bibfield  {journal} {\bibinfo  {journal} {Phys. Rev. B}\ }\textbf {\bibinfo {volume} {85}},\ \bibinfo {pages} {014418} (\bibinfo {year} {2012})}\BibitemShut {NoStop}%
\bibitem [{\citenamefont {Benka}\ \emph {et~al.}(2022)\citenamefont {Benka}, \citenamefont {Bauer}, \citenamefont {Schmakat}, \citenamefont {S\"aubert}, \citenamefont {Seifert}, \citenamefont {Jorba},\ and\ \citenamefont {Pfleiderer}}]{Georg2022}%
  \BibitemOpen
  \bibfield  {author} {\bibinfo {author} {\bibfnamefont {G.}~\bibnamefont {Benka}}, \bibinfo {author} {\bibfnamefont {A.}~\bibnamefont {Bauer}}, \bibinfo {author} {\bibfnamefont {P.}~\bibnamefont {Schmakat}}, \bibinfo {author} {\bibfnamefont {S.}~\bibnamefont {S\"aubert}}, \bibinfo {author} {\bibfnamefont {M.}~\bibnamefont {Seifert}}, \bibinfo {author} {\bibfnamefont {P.}~\bibnamefont {Jorba}},\ and\ \bibinfo {author} {\bibfnamefont {C.}~\bibnamefont {Pfleiderer}},\ }\bibfield  {title} {\bibinfo {title} {Interplay of itinerant magnetism and spin-glass behavior in ${\mathrm{fe}}_{x}\mathrm{Cr}{}_{1\ensuremath{-}x}$},\ }\href {https://doi.org/10.1103/PhysRevMaterials.6.044407} {\bibfield  {journal} {\bibinfo  {journal} {Phys. Rev. Mater.}\ }\textbf {\bibinfo {volume} {6}},\ \bibinfo {pages} {044407} (\bibinfo {year} {2022})}\BibitemShut {NoStop}%
\bibitem [{\citenamefont {Souletie}\ and\ \citenamefont {Tholence}(1985)}]{Souletie1985}%
  \BibitemOpen
  \bibfield  {author} {\bibinfo {author} {\bibfnamefont {J.}~\bibnamefont {Souletie}}\ and\ \bibinfo {author} {\bibfnamefont {J.~L.}\ \bibnamefont {Tholence}},\ }\bibfield  {title} {\bibinfo {title} {Critical slowing down in spin glasses and other glasses: Fulcher versus power law},\ }\href {https://doi.org/10.1103/PhysRevB.32.516} {\bibfield  {journal} {\bibinfo  {journal} {Phys. Rev. B}\ }\textbf {\bibinfo {volume} {32}},\ \bibinfo {pages} {516} (\bibinfo {year} {1985})}\BibitemShut {NoStop}%
\bibitem [{\citenamefont {Küchler}\ \emph {et~al.}(2006)\citenamefont {Küchler}, \citenamefont {Gegenwart}, \citenamefont {Weickert}, \citenamefont {Oeschler}, \citenamefont {Cichorek}, \citenamefont {Nicklas}, \citenamefont {Carocca-Canales}, \citenamefont {Geibel},\ and\ \citenamefont {Steglich}}]{Kuechler2006}%
  \BibitemOpen
  \bibfield  {author} {\bibinfo {author} {\bibfnamefont {R.}~\bibnamefont {Küchler}}, \bibinfo {author} {\bibfnamefont {P.}~\bibnamefont {Gegenwart}}, \bibinfo {author} {\bibfnamefont {F.}~\bibnamefont {Weickert}}, \bibinfo {author} {\bibfnamefont {N.}~\bibnamefont {Oeschler}}, \bibinfo {author} {\bibfnamefont {T.}~\bibnamefont {Cichorek}}, \bibinfo {author} {\bibfnamefont {M.}~\bibnamefont {Nicklas}}, \bibinfo {author} {\bibfnamefont {N.}~\bibnamefont {Carocca-Canales}}, \bibinfo {author} {\bibfnamefont {C.}~\bibnamefont {Geibel}},\ and\ \bibinfo {author} {\bibfnamefont {F.}~\bibnamefont {Steglich}},\ }\bibfield  {title} {\bibinfo {title} {Thermal expansion and grüneisen ratio near quantum critical points},\ }\href@noop {} {\bibfield  {journal} {\bibinfo  {journal} {Physica B: Condensed Matter}\ }\textbf {\bibinfo {volume} {378-380}},\ \bibinfo {pages} {36} (\bibinfo {year} {2006})}\BibitemShut {NoStop}%
\bibitem [{\citenamefont {Westerkamp}\ \emph {et~al.}(2009)\citenamefont {Westerkamp}, \citenamefont {Deppe}, \citenamefont {K\"uchler}, \citenamefont {Brando}, \citenamefont {Geibel}, \citenamefont {Gegenwart}, \citenamefont {Pikul},\ and\ \citenamefont {Steglich}}]{Westerkamp2009}%
  \BibitemOpen
  \bibfield  {author} {\bibinfo {author} {\bibfnamefont {T.}~\bibnamefont {Westerkamp}}, \bibinfo {author} {\bibfnamefont {M.}~\bibnamefont {Deppe}}, \bibinfo {author} {\bibfnamefont {R.}~\bibnamefont {K\"uchler}}, \bibinfo {author} {\bibfnamefont {M.}~\bibnamefont {Brando}}, \bibinfo {author} {\bibfnamefont {C.}~\bibnamefont {Geibel}}, \bibinfo {author} {\bibfnamefont {P.}~\bibnamefont {Gegenwart}}, \bibinfo {author} {\bibfnamefont {A.~P.}\ \bibnamefont {Pikul}},\ and\ \bibinfo {author} {\bibfnamefont {F.}~\bibnamefont {Steglich}},\ }\bibfield  {title} {\bibinfo {title} {Kondo-cluster-glass state near a ferromagnetic quantum phase transition},\ }\href {https://doi.org/10.1103/PhysRevLett.102.206404} {\bibfield  {journal} {\bibinfo  {journal} {Phys. Rev. Lett.}\ }\textbf {\bibinfo {volume} {102}},\ \bibinfo {pages} {206404} (\bibinfo {year} {2009})}\BibitemShut {NoStop}%
\bibitem [{\citenamefont {Kawasaki}\ \emph {et~al.}(2009)\citenamefont {Kawasaki}, \citenamefont {Nishikawa}, \citenamefont {Hidaka}, \citenamefont {Yanagisawa}, \citenamefont {Tenya}, \citenamefont {Yokoyama},\ and\ \citenamefont {Amitsuka}}]{Kawasaki2009}%
  \BibitemOpen
  \bibfield  {author} {\bibinfo {author} {\bibfnamefont {I.}~\bibnamefont {Kawasaki}}, \bibinfo {author} {\bibfnamefont {D.}~\bibnamefont {Nishikawa}}, \bibinfo {author} {\bibfnamefont {H.}~\bibnamefont {Hidaka}}, \bibinfo {author} {\bibfnamefont {T.}~\bibnamefont {Yanagisawa}}, \bibinfo {author} {\bibfnamefont {K.}~\bibnamefont {Tenya}}, \bibinfo {author} {\bibfnamefont {M.}~\bibnamefont {Yokoyama}},\ and\ \bibinfo {author} {\bibfnamefont {H.}~\bibnamefont {Amitsuka}},\ }\bibfield  {title} {\bibinfo {title} {Magnetic properties around quantum critical point of cept1-xrhx},\ }\href {https://doi.org/https://doi.org/10.1016/j.physb.2009.07.139} {\bibfield  {journal} {\bibinfo  {journal} {Physica B: Condensed Matter}\ }\textbf {\bibinfo {volume} {404}},\ \bibinfo {pages} {2908} (\bibinfo {year} {2009})}\BibitemShut {NoStop}%
\bibitem [{\citenamefont {Pikul}\ and\ \citenamefont {Kaczorowski}(2012)}]{Pikul2012}%
  \BibitemOpen
  \bibfield  {author} {\bibinfo {author} {\bibfnamefont {A.~P.}\ \bibnamefont {Pikul}}\ and\ \bibinfo {author} {\bibfnamefont {D.}~\bibnamefont {Kaczorowski}},\ }\bibfield  {title} {\bibinfo {title} {Search for quantum criticality in a ferromagnetic system uni${}_{1\ensuremath{-}x}$co${}_{x}$si${}_{2}$},\ }\href {https://doi.org/10.1103/PhysRevB.85.045113} {\bibfield  {journal} {\bibinfo  {journal} {Phys. Rev. B}\ }\textbf {\bibinfo {volume} {85}},\ \bibinfo {pages} {045113} (\bibinfo {year} {2012})}\BibitemShut {NoStop}%
\bibitem [{\citenamefont {Pikul}(2012)}]{Pikul20122}%
  \BibitemOpen
  \bibfield  {author} {\bibinfo {author} {\bibfnamefont {A.~P.}\ \bibnamefont {Pikul}},\ }\bibfield  {title} {\bibinfo {title} {The influence of magnetic sublattice dilution on magnetic order in cenige3 and unisi2},\ }\href {https://doi.org/10.1088/0953-8984/24/27/276003} {\bibfield  {journal} {\bibinfo  {journal} {Journal of Physics: Condensed Matter}\ }\textbf {\bibinfo {volume} {24}},\ \bibinfo {pages} {276003} (\bibinfo {year} {2012})}\BibitemShut {NoStop}%
\end{thebibliography}
%

\end{document}